\documentclass[MSNbibl,nameyear,rotating,dvips]{arxstspdf}
\usepackage{graphicx}
\usepackage{flushend}
\usepackage{stfloats}

\volume{27}
\issue{2}
\pubyear{2012}
\firstpage{247}
\lastpage{277}
\doi{10.1214/11-STS370}

\makeatletter

\def\citep#1{(\citeauthor{#1}, \citeyear{#1})}

\setattribute{keyword}{width}{330pt}

\makeatother

\begin{document}
\begin{frontmatter}

\title{Estimators of Fractal Dimension:
Assessing the Roughness of Time Series and Spatial Data}
\runtitle{Estimators of Fractal Dimension}

\begin{aug}
\author[a]{\fnms{Tilmann} \snm{Gneiting}\corref{}\ead[label=e1]{t.gneiting@uni-heidelberg.de}},
\author[b]{\fnms{Hana} \snm{\v{S}ev\v{c}\'{\i}kov\'{a}}\ead[label=e2]{hanas@uw.edu}}
\and
\author[c]{\fnms{Donald B.} \snm{Percival}\ead[label=e3]{dbp@apl.washington.edu}}
\runauthor{T. Gneiting, H. \v{S}ev\v{c}\'{\i}kov\'{a}, and D. B. Percival}
\pdfauthor{T. Gneiting, H. Sevcikova, D. B. Percival}

\affiliation{Universit\"at Heidelberg, University of Washington, and
University of Washington}

\address[a]{Tilmann Gneiting is Professor,
Institut f\"ur Angewandte Mathematik, Universit\"at Heidelberg, Im
Neuenheimer Feld 294, 69120 Heidelberg, Germany
\printead{e1}.}
\address[b]{Hana \v{S}ev\v{c}\'{\i}kov\'{a} is Senior Research
Scientist, Center for Statistics and the Social Sciences, University of
Washington, Box 354322, Seattle, Washington 98195-4322, USA \printead{e2}.}
\address[c]{Donald B. Percival is Principal Mathematician, Applied
Physics Laboratory; Professor,
Department of Statistics, University of
Washington, Box 355640, Seattle, Washington 98195-5640, USA \printead{e3}.}

\end{aug}

%
\begin{abstract}
The fractal or Hausdorff dimension is a measure of roughness (or
smoothness) for time series and spatial data. The graph of a~smooth,
differentiable surface indexed in $\mathbb{R}^d$ has topological and
fractal dimension $d$. If the surface is nondifferentiable and
rough, the fractal dimension takes values between the topological
dimension, $d$, and $d+1$. We review and assess estimators of fractal
dimension by their large sample behavior under infill asymptotics, in
extensive finite sample simulation studies, and in a data example on
arctic sea-ice profiles. For time series or line transect data,
box-count, Hall--Wood, semi-periodogram, discrete cosine transform and
wavelet estimators are studied along with variation estimators with
power indices 2 (variogram) and 1 (madogram), all implemented in the
\texttt{R} package \texttt{fractaldim}. Considering both efficiency and
robustness, we recommend the use of the madogram estimator, which can
be interpreted as a statistically more efficient version of the
Hall--Wood estimator. For two-dimensional lattice data, we propose
robust transect estimators that use the median of variation estimates
along rows and columns. Generally, the link between power variations
of index $p > 0$ for stochastic processes, and the Hausdorff dimension
of their sample paths, appears to be particularly robust and inclusive
when $p = 1$.
\end{abstract}

%

\end{frontmatter}

\begin{flushright} \textit{Lies, damn lies, and dimension estimates}

 Lenny Smith [(\citeyear{Smith07}), p.~115]
\end{flushright}
\section{Introduction}

Fractal-based analyses of time series, transects, and natural or
man-made surfaces have found extensive applications in almost all
scientific disciplines (Mandelbrot, \citeyear{mandelbrot82}). While much of the
literature ties fractal properties to statistical self-similarity, no
such link is necessary. Rather, we adopt the argument of Bruno and Raspa (\citeyear{BR89}),
Davies and Hall (\citeyear{DH99}) and Gneiting and Schlather (\citeyear{GnSch04}) that the fractal or Hausdorff dimension
quantifies the roughness or smoothness of time series and spatial
data in the limit as the observational scale becomes infinitesimally
fine. In practice, measurements can only be taken at a finite range
of scales, and usable estimates of the fractal dimension depend on the
availability of observations at sufficiently fine temporal or spatial
resolution (Malcai et~al., \citeyear{MLBA97}; Halley et~al.,
\citeyear{halley04}).

We follow common practice in defining the fractal dimension of a point
set $X \subset\mathbb{R}^d$ to be the classical Hausdorff dimension
(Hausdorff, \citeyear{hausdorff19}; Falconer, \citeyear{falconer90}). For $\varepsilon> 0$, an
$\varepsilon$-cover of $X$ is a finite or countable collection $\{ B_i
\dvtx
i = 1, 2, \ldots\}$ of balls $B_i \subset\mathbb{R}^d$ of diameter
$|B_i|$ less than or equal to $\varepsilon$ that covers $X$. With
\begin{eqnarray*}
H^\delta(X) &=& \lim_{\varepsilon\to0}   \inf\Biggl\{ \sum
_{i=1}^\infty
|B_i|^\delta\dvtx \{ B_i \dvtx i = 1, 2, \ldots\}
\\
&&\hspace*{77pt}\mbox{is an $\varepsilon
$-cover of } X \Biggr\}
\end{eqnarray*}
denoting the $\delta$-dimensional Hausdorff measure of $X$, there
exists a unique nonnegative value $D$ such that $H^\delta(X) = \infty$
if $\delta< D$ and $H^\delta(X) = 0$ if $\delta> D$. This value $D$
is the Hausdorff dimension of the point set~$X$. Under weak
regularity conditions, the Hausdorff dimension coincides with the
box-count dimension,
%
\begin{equation} \label{eqbcfd}
D_{\rm BC} = \lim_{\varepsilon\to0} \frac{\log N(\varepsilon)}{\log
(1/\varepsilon)},
\end{equation}
where $N(\varepsilon)$ denotes the smallest number of cubes of width
$\varepsilon$ in $\mathbb{R}^d$ which can cover $X$, and also with other
natural and/or time-honored notions of dimension (Falconer, \citeyear{falconer90}).

\begin{table*}[t]
\tablewidth=350pt
\caption{Some parametric classes of variograms and covariance
functions for a Gaussian process $\{ X_t \dvtx t \in\mathbb{R}^d \}$. The
covariance functions have been normalized such that $\sigma(0) = 1$.
Here, $\alpha$ is the fractal index, $c > 0$ is a range parameter,
and $K_\nu$ is a modified Bessel function of the second kind of
order $\nu$. The Mat\'ern and Dagum families allow for less
restrictive assumptions on the parameters than stated here}
\label{tabcov}
\begin{tabular*}{350pt}{@{\extracolsep{4in minus 4in}}lll@{}}
\hline
{\textbf{Class}}& {\textbf{Variogram or covariance}}&{\textbf{Parameters}} \\
\hline
Fractional Brownian motion & $\gamma_2(t) = |c \hspace
{0.3mm}t|^\alpha$
& $\alpha\in(0,2]$ \\
Mat\'ern & $\sigma(t) = \frac{2^{(\alpha/2)-1}}{\Gamma(\alpha/2)}
|c  t|^{\alpha/2}  K_{\alpha/2}(|c
 t|)$
& $\alpha\in(0,2]$ \\
Powered exponential & $\sigma(t) = \exp(-|c  t|^\alpha
)$
& $\alpha\in(0,2]$ \\
Cauchy & $\sigma(t) = (1+|c  t|^\alpha)^{-\tau/\alpha
}$
& $\alpha\in(0,2];   \tau> 0$ \\
Dagum & $\sigma(t) = 1 - ( \frac{|c  t|^\tau}{1
+ |c
t|^\tau} )^{\alpha/\tau}$
& $\tau\in(0,2];   \alpha\in(0,\tau)$ \\
\hline
\end{tabular*}
\end{table*}

In this paper we restrict attention to the case in which the point set
\[
X = \{   (t, X_t ) \in\mathbb{R}^d  \times \mathbb{R}\dvtx
 t \in{\mathrm T}\subset\mathbb{R}^d   \} \subset
\mathbb{R}^{d+1}
\]
is the graph of time series or spatial data observed at a finite set
${\mathrm T}\subset\mathbb{R}^d$ of typically regularly spaced times or
locations. The fractal dimension then refers to the properties of the
curve ($d = 1$) or surface ($d \geq2$) that arises in the continuum
limit as the data are observed at an infinitesimally dense subset of
the temporal or spatial domain, which without loss of generality can
be assumed to be the unit interval or unit cube. In time series
analysis and spatial statistics, this limiting scenario is referred to
as infill asymptotics (Hall and Wood, \citeyear{hallwood93}; Dahlhaus,
\citeyear{dahlhaus97}; Stein, \citeyear{stein99}).\vadjust{\goodbreak}

If the  limit curve or limit surface is smooth and differentiable, its
fractal dimension, $D$, equals its topological dimension, $d$. For a
rough and nondifferentiable curve or surface, the fractal dimension
may exceed the topological dimension. For example, suppose that $\{
X_t \dvtx t \in\mathbb{R}^d \}$ is a Gaussian process with stationary
increments, whose variogram or structure function,
%
\begin{equation} \label{eqdefvariogram}
\gamma_2(t) = \tfrac{1}{2}   {\mathbb{E}}( X_u - X_{u+t}
)^2  ,
\end{equation}
satisfies
%
\begin{equation} \label{eqvariogram}
\gamma_2(t) = |c_2  t|^\alpha+ \mathcal{O}(|t|^{\alpha
+\beta})
\quad\mbox{as }  t \to0,
\end{equation}
where $\alpha\in(0,2]$, $\beta\geq0$, and $c_2 > 0$, and $| \cdot
|$ denotes the Euclidean norm. Then the graph of a sample path has
fractal dimension
%
\begin{equation} \label{eqDalpha}
D = d + 1 - \frac{\alpha}{2}
\end{equation}
almost surely (Orey, \citeyear{orey70}; Adler \citeyear{adler81}). This relationship links the
fractal dimension of the sample paths to the behavior of the variogram
or structure function at the coordinate origin, and can be extended to
broad classes of potentially anisotropic and nonstationary processes,
as well as some non-Gaussian processes (Adler, \citeyear{adler81}; Hall and Roy, \citeyear{HR94}; Xue and Xiao, \citeyear{XX09}).
It allows us to think of fractal dimension as a second-order property
of a Gaussian stochastic process, in addition to being a roughness
measure for a realized curve or surface. Accordingly, we refer to the
index $\alpha$ in the asymptotic relationship (\ref{eqvariogram}) as
the fractal index.

Table \ref{tabcov} provides examples of Gaussian processes that
exhibit this asymptotic behavior. Fractional\break Brownian motion is a
nonstationary, statistically self-similar process that is defined in
terms of the variogram (Mandelbrot and Van~Ness, \citeyear{MvN68}). The other entries in the table
refer to stationary processes with covariance function $\sigma(t) =
{\rm cov} (X_u,X_{t+u})$, which relates to the variogram as
\[
\gamma_2(t) = \sigma(0) - \sigma(t), \quad t \in\mathbb{R}^d.
\]
Key examples include the Mat\'ern family (Mat{\'e}rn, \citeyear{Matern60}; Guttorp and Gneiting, \citeyear{gg06}), used
by Goff and Jordan (\citeyear{goffjordan88}) to parameterize the fractal dimension of
oceanic features; the Cauchy class, introduced by Gneiting and Schlather (\citeyear{GnSch04}) to
illustrate local and global properties of random functions; and the
Dagum family (Berg, Mateu and Porcu, \citeyear{BMP08}). In simulation settings, the powered
exponential class (Yaglom, \citeyear{yaglom}) is a convenient default example. The
exponent $\beta$ in the asymptotic relationship (\ref{eqvariogram})
equals $\beta= 2 - \alpha$ for\vadjust{\goodbreak} the Mat\'ern class, $\beta= \alpha$
for the powered exponential and Cauchy families, and $\beta= \tau$
for the Dagum family. Generally, the smaller the value of~$\beta$,
the harder the estimation of the fractal index,~$\alpha$, and the more
pronounced the finite sample bias of estimators of the fractal index
or fractal dimension.\looseness=-1

\begin{figure*}

\includegraphics{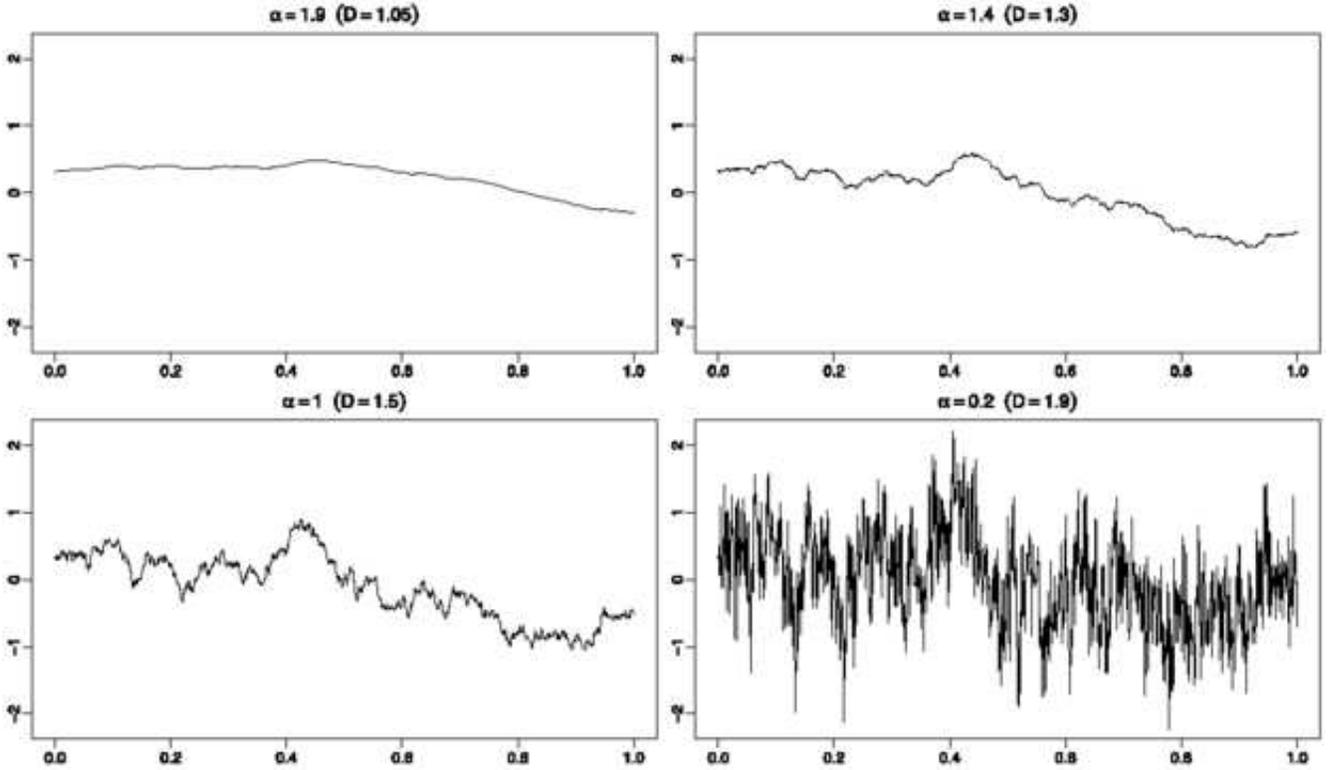}

\caption{Sample paths of stationary Gaussian processes with powered
exponential covariance function, $\sigma(t) = \exp(-|t|^\alpha)$,
and fractal index, $\alpha$, equal to 1.9, 1.4, 1.0 and 0.2. The
corresponding values of the fractal dimension, $D$, are 1.05, 1.3,
1.5 and 1.9, respectively. The simulation domain is a grid over the
unit interval with spacing $1/1{,}024$.}\label{fig1d}
\end{figure*}

\begin{figure*}

\includegraphics{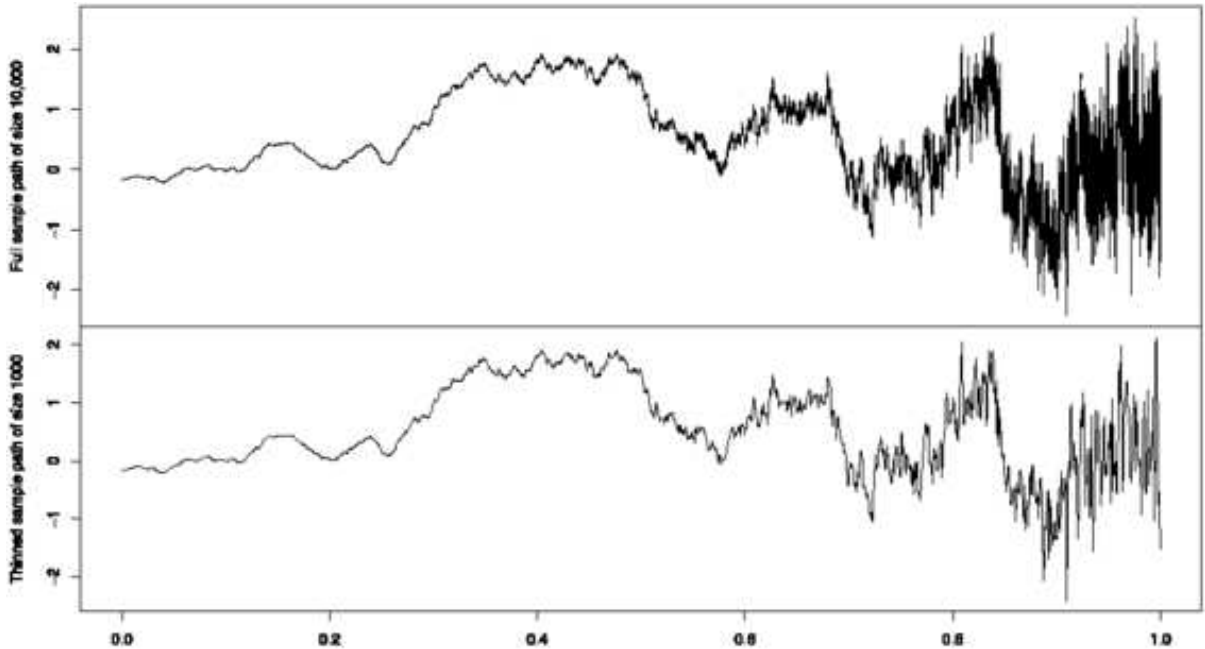}

\caption{A sample path from the nonstationary Gaussian Mat\'ern
process of Anderes and Stein (\protect\citeyear{anderesstein10}),
where the fractal dimension, $D$,
grows linearly from $D = 1$ at time $t = 0$ to $D = 2$ at time $t =
1$. To illustrate the visual effects of the temporal resolution,
both the original sample path with grid spacing $1/10{,}000$ (top
panel) and an equidistantly thinned version with grid spacing
$1/1{,}000$ (bottom panel) are shown. The nonstationary covariance
is given by equation~(10) of Anderes and Stein (\protect\citeyear{anderesstein10})
with $\sigma^2 = 1$, $\rho= 1/2$, and linearly varying local smoothness parameter,
$\nu_t = 1 - t$.}
\label{figwm-nonstationary}
\end{figure*}

As an illustration for time series or line transect data,
Figure~\ref{fig1d} displays Gaussian sample paths from the powered
exponential family with the fractal index, $\alpha$, ranging from 1.9
to 0.2, and the fractal dimension, $D$, extending from 1.05 to 1.9.
The small\-er~$\alpha$, the rougher the sample path, and the larger the
fractal dimension, with the lower limit, $D = 1$, being associated
with a smooth curve, and the upper limit, $D = 2$, corresponding to a
space-filling, exceedingly rough graph.
Figure~\ref{figwm-nonstationary} shows a realization from the
nonstationary Gaussian Mat\'ern model of Anderes and Stein (\citeyear{anderesstein10}), in a
case in which the fractal index and the fractal dimension vary
linearly along the unit interval. To illustrate the visual effects of
the measurement scale, both the original sample path of size 10,000
and an equidistantly thinned version of size 1,000 are shown.

\begin{figure*}

\includegraphics{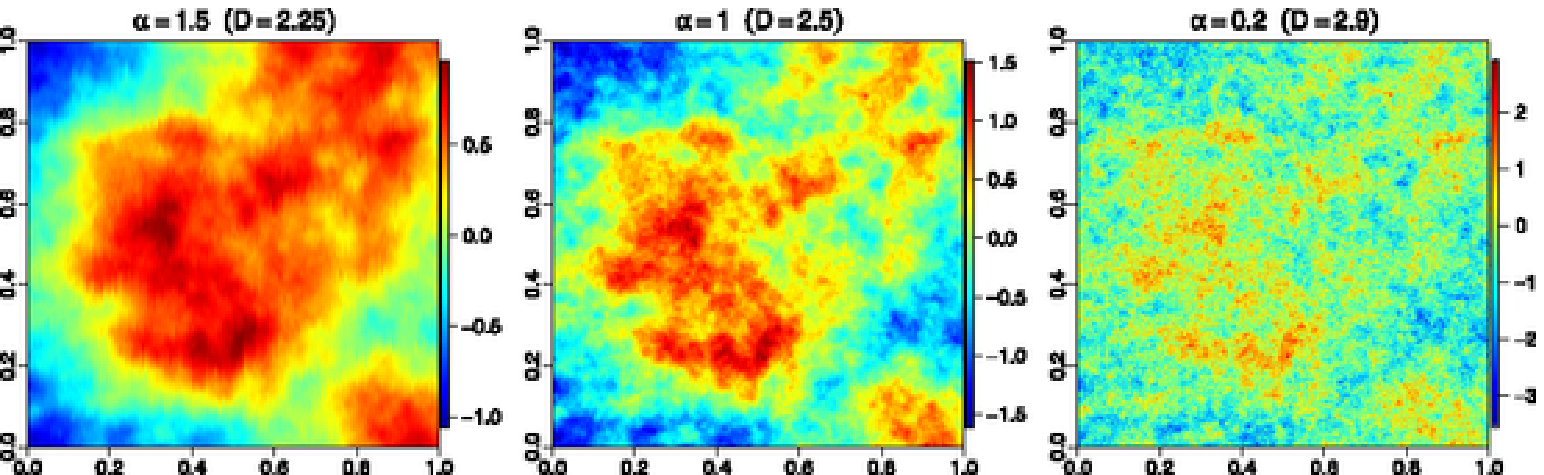}

\caption{Sample surfaces of stationary spatial Gaussian processes with
the powered exponential covariance function, $\sigma(t) =
\exp(-|t|^\alpha)$, and fractal index, $\alpha$, equal to 1.5, 1.0
and 0.2. The corresponding values of the fractal dimension, $D$,
are 2.25, 2.5 and 2.9. The simulation domain is a grid on the unit
square in $\mathbb{R}^2$ with spacing $1/512$ along each coordinate.
}\label{fig2d}
\end{figure*}

Turning to spatial data, Figure \ref{fig2d} shows Gaussian sample
surfaces from the powered exponential class with the fractal index,
$\alpha$, being equal to 1.5, 1.0 and 0.2. The surfaces are
increasingly rough with the fractal dimension, $D$, being equal to
2.25, 2.5 and 2.9, respectively.

A wealth of applications requires the characterization of the roughness
or smoothness of time series, line transect or spatial data, with
Burrough (\citeyear{burrough81}) and Malcai et~al. (\citeyear{MLBA97}) summarizing an impressive range
of experimental results. For example, fractal dimensions have been
studied for geographic profiles and surfaces, such as the underside of
sea ice (Rothrock and Thorndike, \citeyear{rothrock80}), the topography of the sea floor
(Goff and Jordan, \citeyear{goffjordan88}),\vadjust{\goodbreak} Martial surface (Orosei et~al., \citeyear{oroseietal03}) and
terrestrial features (Weissel, Pratson and Malinverno, \citeyear{weisseletal94}; Turcotte, \citeyear{Turcotte92}; Gagnon, Lovejoy and Schertzer, \citeyear{gagnonetal06}).
Further references can be found in Molz, Rajaram and Lu (\citeyear{molzetal04}) for applications in subsurface hydrology and in Halley et~al. (\citeyear{halley04}) for applications in ecology.
Not surprisingly then, estimators for the fractal
dimension have been proposed and widely used in various literatures,
including physics, engineering, the earth sciences, and statistics,
with the works of Burrough (\citeyear{burrough81}), Goff and Jordan (\citeyear{goffjordan88}), Bruno and Raspa (\citeyear{BR89}),
Dubuc et~al. (\citeyear{dubuc89a,dubuc89}), Jakeman and Jordan (\citeyear{JJ90}), Theiler (\citeyear{theiler90}),
Klinkenberg and Goodchild (\citeyear{klinkenbergetal92}), Schepers, van Beek and Bassingthwaighte (\citeyear{schepersetal92}), Hall and Wood (\citeyear{hallwood93}),
Constantine and Hall (\citeyear{consthall94}), Kent and Wood (\citeyear{kentwood97}), Davies and Hall (\citeyear{DH99}), Chan and Wood (\citeyear{chanwood00}),
Zhu and Stein (\citeyear{zhustein02}), Chan and Wood (\citeyear{chanwood04}) and Bez and Ber\-trand (\citeyear{BB10}) being examples.
Our objectives in this paper are to survey the literature across
disciplines, review and assess the various types of estimators, and
provide recommendations for practitioners, along with novel directions
for theoretical work.\looseness=-1

The remainder of the paper is organized as follows. In Section
\ref{secestimators1d} we describe estimators of fractal dimension for
time series and line transect data, including box-count, Hall--Wood,
variogram, madogram,~pow\-er variation, semi-periodogram and wavelet-based
techniques. Then in Section \ref{secperformance1d} we assess
and compare the estimators. Considering both efficiency and
robustness, we concur with Bruno and Raspa (\citeyear{BR89}) and Bez and Ber\-trand (\citeyear{BB10}) and recommend
the use of the madogram estimator, that is, the variation estimator
with power index $p = 1$, which can be interpreted as a statistically
efficient version of the Hall--Wood estimator. An underlying
motivation is that for\vadjust{\goodbreak} an intrinsically stationary process with
variogram of order $p > 0$ of the form
\begin{eqnarray} \label{eqgammap}
\quad \gamma_{ p}(t)
&=& \tfrac{1}{2}   {\mathbb{E}}| X_u - X_{u+t} |^p\nonumber
\\[-8pt]\\[-8pt]
&=& |c_{ p}  t|^{\alpha p/2}+
\mathcal{O}\bigl(|t|^{(\alpha+\beta)
 p/2}\bigr)
\quad\mbox{as }  t \to0,\nonumber
\end{eqnarray}
the relationship (\ref{eqDalpha}) between the fractal index,
$\alpha$, and the fractal dimension, $D$, appears to be more robust
and inclusive when $p = 1$, as compared to the default case, in which
$p = 2$.

Section \ref{secestimators2d} discusses ways in which estimators for
the time series or line transect case can be adapted to spatial data
observed over a regular lattice in $\mathbb{R}^2$, and Section
\ref{secperformance2d} evaluates these proposals. Our preferred
estimators in this setting are simple, robust transect estimators that
employ the median of variation estimates with power index $p = 1$
along individual rows and columns. A data example on arctic sea-ice
profiles is given in Section \ref{secdata}. The paper ends in
Section \ref{secdiscussion}, where we make a call for new directions
in theoretical and methodological work that addresses both
probabilists and statisticians. Furthermore, we hint at inference for
nonstationary or multifractional processes, where the fractal
dimension of a sample path may vary temporally or spatially. All
computations in the paper use the \texttt{fractaldim} package
({\v{S}}ev{\v{c}}{\'{\i}}kov{\'a}, Gneiting and Percival, \citeyear{sevc10}), which implements our proposals in \texttt{R}
(Ihaka and Gentleman, \citeyear{IG96}).

\section{Estimating the Fractal Dimension of Time Series and Line Transect Data} \label{secestimators1d}

Spurred and inspired by the now classical essay of
Mandelbrot (\citeyear{mandelbrot82}), a large number of methods have been developed for
estimating fractal dimension.\vadjust{\goodbreak} By the early 1990s a sizable, mostly
heuristic literature on the estimation of fractal dimension for time
series and line transect data had accumulated in the physical,
engineering and earth sciences, where various reviews are available
(Dubuc et~al., \citeyear{dubuc89a}; Klinkenberg and Goodchild, \citeyear{klinkenbergetal92}; Schepers, van Beek and
Bassingthwaighte, \citeyear{schepersetal92}; Gallant et~al., \citeyear{gallantetal94};
Klinkenberg, \citeyear{klink94}; Schmittbuhl, Vilotte and Roux, \citeyear{schmittbuhletal95}).
These developments prompted the
statistical community to introduce new methodology, along with
asymptotic theory for box-count (Hall and Wood, \citeyear{hallwood93}), variogram
(Constantine and Hall, \citeyear{consthall94}; Kent and Wood, \citeyear{kentwood97}), level crossing
(Feuer\-verger, Hall and Wood, \citeyear{feuerhallwood94}) and spectral (Chan, Hall and Poskitt, \citeyear{chanhall95}) estimators,
among others.

Essentially all methods follow a common scheme, in that:
\begin{longlist}[(a)]
\item[(a)]
a certain numerical property, say $Q$, of the time series or line
transect data is computed as a~function of ``scale,'' say $\varepsilon$;
\item[(b)]
an asymptotic power law $Q(\varepsilon) \propto\varepsilon^b$ as the scale
$\varepsilon\to0$ becomes infinitesimally small is derived or
postulated; where
\item[(c)]
the scaling exponent, $b$, is a linear function of the fractal
dimension, $D$;
\item[(d)]
and thus $D$ is estimated by linear regression of $\log Q(\varepsilon)$
on $\log\varepsilon$, with emphasis on the smallest observed values of
the scale $\varepsilon$.
\end{longlist}
Table \ref{tabfdmethods} shows the data property, the measure of
scale and the scaling law for various methods. For techniques working
in the spectral domain, the scaling law applies as the frequency grows
to infinity, equivalent to the scale becoming infinitesimally small in
the time domain.

\begin{table*}[t]
\tablewidth=380pt
\caption{Some extant methods for estimating the fractal dimension of
time series and line transect data}\label{tabfdmethods}
\begin{tabular*}{380pt}{@{\extracolsep{4in minus 4in}}lllll@{}}
\hline
{\textbf{Method}} & {\textbf{Property}} &{\textbf{Scale}} & {\textbf{Scaling law}} & {\textbf{Regime}} \\
\hline
Box-count & $N(\varepsilon)$: number of boxes
& $\varepsilon$: box width
& $N(\varepsilon) \propto\varepsilon^{-D}$
& $\varepsilon\to0$  \\
Divider & $L(\varepsilon)$: length of curve
& $\varepsilon$: step size
& $L(\varepsilon) \propto\varepsilon^{-1-D}$
& $\varepsilon\to0$  \\
Level crossing & $M(h)$: number of crossings
& $h$: bandwidth
& $M(h) \propto h^{1-D}$
& $h \to0$  \\
Variogram & $\gamma_2(t)$: variogram
& $t$: lag
& $\gamma_2(t) \propto t^{4-2D}$
& $t \to0$  \\
Madogram & $\gamma_1(t)$: madogram
& $t$: lag
& $\gamma_1(t) \propto t^{2-D}$
& $t \to0$  \\
Spectral & $f(\omega)$: spectral density
& $\omega$: frequency
& $f(\omega) \propto\omega^{2D-5}$
& $\omega\to\infty$ \\
Wavelet & $\nu^2(\tau)$: wavelet variance
& $\tau$: scale
& $\nu^2(\tau) \propto\tau^{4-2D}$
& $\tau\to0$
\\
\hline
\end{tabular*}
\end{table*}

In the balance of this section, we describe the most popular
estimators of fractal dimension in the equally spaced time series or
line transect setting. Without loss of generality, we may assume that
the observation domain is the unit interval. In the case of $n_s = n
+ 1$ equally spaced observations, the data graph is the point set
%
\begin{equation} \label{eqtsdata}
\biggl\{ (t, X_t)  \dvtx  t =
\frac{i}{n},   i = 0, 1,
\ldots, n \biggr\} \subset\mathbb{R}^2.
\end{equation}
The relevant asymptotic regime then is infill asymptotics, in which
the number of observations grows to infinity, whereas the underlying
domain, namely the unit interval, remains fixed. For convenience in
what follows, we refer to both $n$ and $n_s$ as the sample size.

\subsection{Box-Count Estimator} \label{secbox-count}

The popular box-count estimator is motivated by the scaling law
(\ref{eqbcfd}) that defines the box-count dimension. The basic idea
is simple, in that the time series graph is initially covered by a
single box. The box is divided into four quadrants, and the number of
cells required to cover the curve is counted. Then each subsequent
quadrant is divided into four subquadrants, and one continues doing so
until the box width equals the resolution of the data, keeping track
of the number of quadrants required to cover the graph at each step.
If $N(\varepsilon)$ denotes the number of boxes required at width or
scale $\varepsilon$, the box-count estimator equals the slope in an
ordinary least squares regression fit of $\log N(\varepsilon)$ on $\log
\varepsilon$. Similarly to Mandelbrot's (1967) divider technique, the
method can be used to quantify the fractal dimension of any planar
point set, rather than just equally spaced time series or line transect
data.

%
\begin{figure*}[t]

\includegraphics{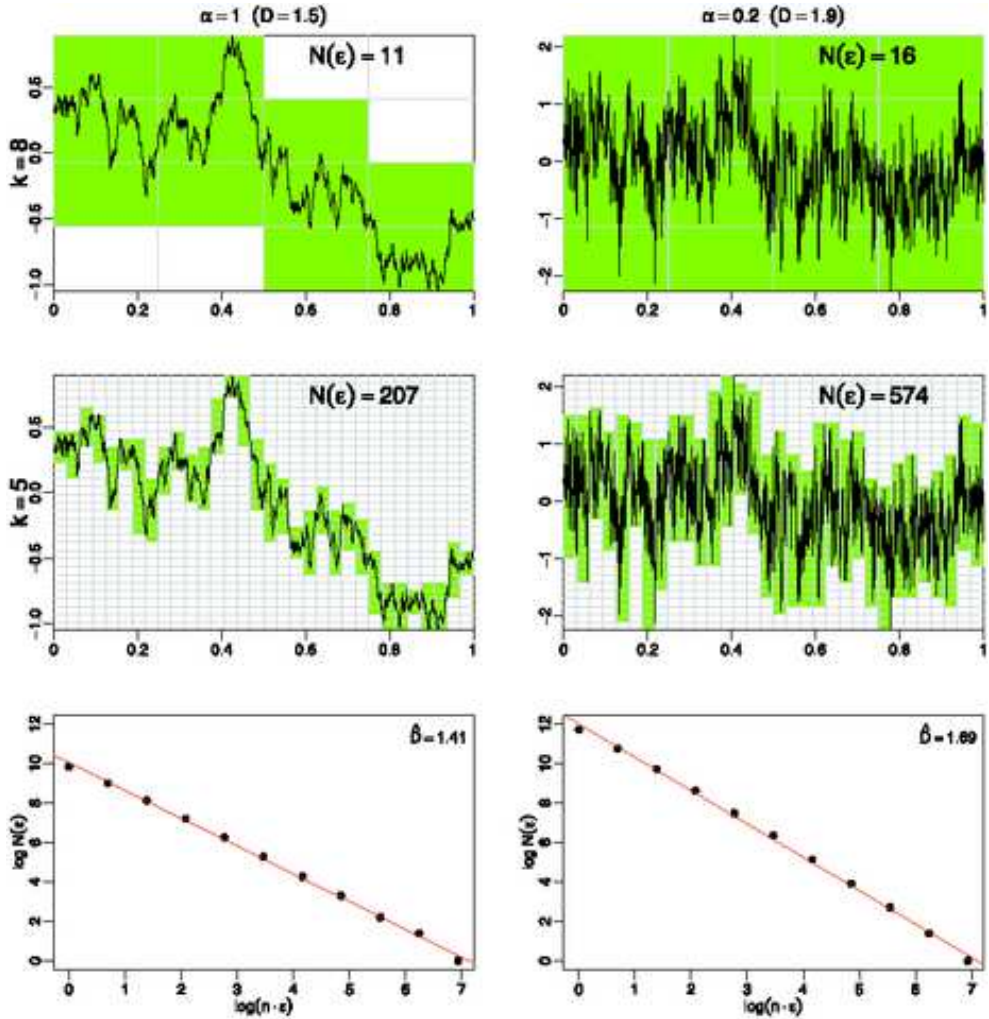}

\caption{Illustration of the box-count algorithm and naive box-count
estimates for the two datasets in the lower row of
Figure~\protect\ref{fig1d}. See the text for
details.} \label{figbox-count-howto}\vspace*{3pt}
\end{figure*}

\begin{figure*}[b]
\vspace*{3pt}
\includegraphics{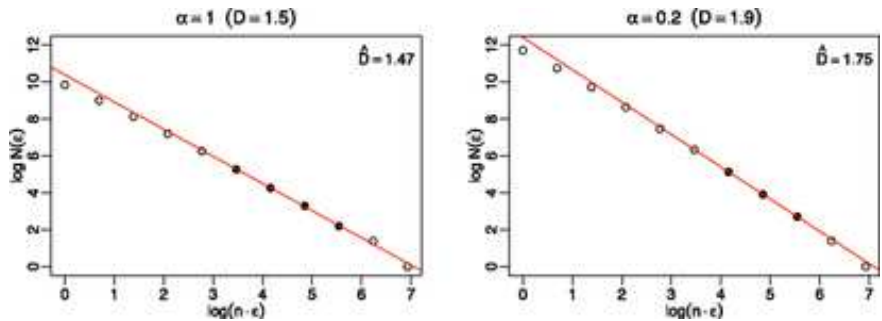}

\caption{Log-log regression for the standard version of the box-count
estimator and the datasets in the lower row of Figure~\protect\ref{fig1d}.
Only the points marked with filled circles are used when fitting the
regression line.} \label{figbox-count}
\end{figure*}
%

In our setting of a data graph of the form (\ref{eqtsdata}), where,
for simplicity, we assume that $n = 2^K$ is a~pow\-er of 2, the
box-count algorithm can be summarized as follows. Let $u = \max_{0
\leq j \leq n} X_{j/n} - \min_{0 \leq j \leq n} X_{j/n}$ denote the
range of the data. Consider scales $\varepsilon_k = 2^{k-K}$ where $k =
0, 1, \ldots, K$. At the largest scale,\vadjust{\goodbreak} $\varepsilon_K = 1$, the graph
(\ref{eqtsdata}) can be covered by a single box of width $1$ and
height $u$, which we now call the bounding box. At scale $\varepsilon_k$
the bounding box can be tiled by $4^{K-k}$ boxes of width $2^{k-K}$
and height $u  2^{k-K}$ each, and we define $N(\varepsilon
_k)$ to be
the number of such boxes that intersect the linearly interpolated data
graph. Figure \ref{figbox-count-howto} provides an illustration on
two of the datasets in Figure \ref{fig1d}, for which $n = 1024$
and $K = 10$. For example, the upper left plot considers $k = 8$,
where $\varepsilon_8 = 2^{-2}$ and $N(\varepsilon_8) = 11$, and the middle
left plot looks at $k = 5$, where $\varepsilon_5 = 2^{-5}$ and
$N(\varepsilon_5) = 207$. The naive box-count estimator then uses the
slope in an ordinary least squares regression fit of $\log
N(\varepsilon)$ on $\log\varepsilon$, that is,
\[
\widehat{D}_{\mathrm BC}= - \Biggl\{ \sum_{k=0}^K (s_k - \bar{s})
\log N(\varepsilon_k)
\Biggr\}
\Biggl\{ \sum_{k=0}^K (s_k - \bar{s})^2 \Biggr\}^{-1},
\]
where $s_k = \log\varepsilon_k$ and $\bar{s}$ is the mean of $s_0, s_1,
\ldots, s_K$. In our illustrating example, this leads to the
estimates shown in the lower row of Figure \ref{figbox-count-howto}.

Several authors identified problems with the naive estimator that
includes all scales in the regression fit of $\log N(\varepsilon)$ on
$\log\varepsilon$, and proposed modifications that address these issues
(Dubuc et~al., \citeyear{dubuc89a}; Liebovitch and Toth, \citeyear{liebovitch89}; Block, von Bloh and Schellnhuber, \citeyear{blocketal90};
Taylor and Taylor, \citeyear{taylor91}). Indeed, one
always has $N(\varepsilon_0) \geq n$ and $N(\varepsilon_K) = 1$, which
suggests that the counts at both the smallest and the largest scales
ought to be discarded. Liebovitch and Toth (\citeyear{liebovitch89}) proposed to exclude the
smallest scales $\varepsilon_k$ for which $N(\varepsilon_k) > n/5$, as well
as the two largest scales, from the regression fit. We adopt this
proposal in our standard version of the box-count estimator, as
illustrated in Figure~\ref{figbox-count}. The restriction on the
scales improves the statistical and computational efficiency of the
estimator. However, it is in the limit as $\varepsilon\to0$ that the
underlying scaling law (\ref{eqbcfd}) operates, and thus it is
unfortunate that information at the very smallest scales is discarded.\vadjust{\goodbreak}

A natural variant of the box-count estimator uses scales $\varepsilon_l =
l/n$, rather than scales $\varepsilon_k = 2^k/n$ at the powers of 2 only.
In the next section we discuss a~related estimator that takes up this
idea, addresses the aforementioned limitations, and is tailored to
time series data of the form (\ref{eqtsdata}).

\begin{figure*}[t]

\includegraphics{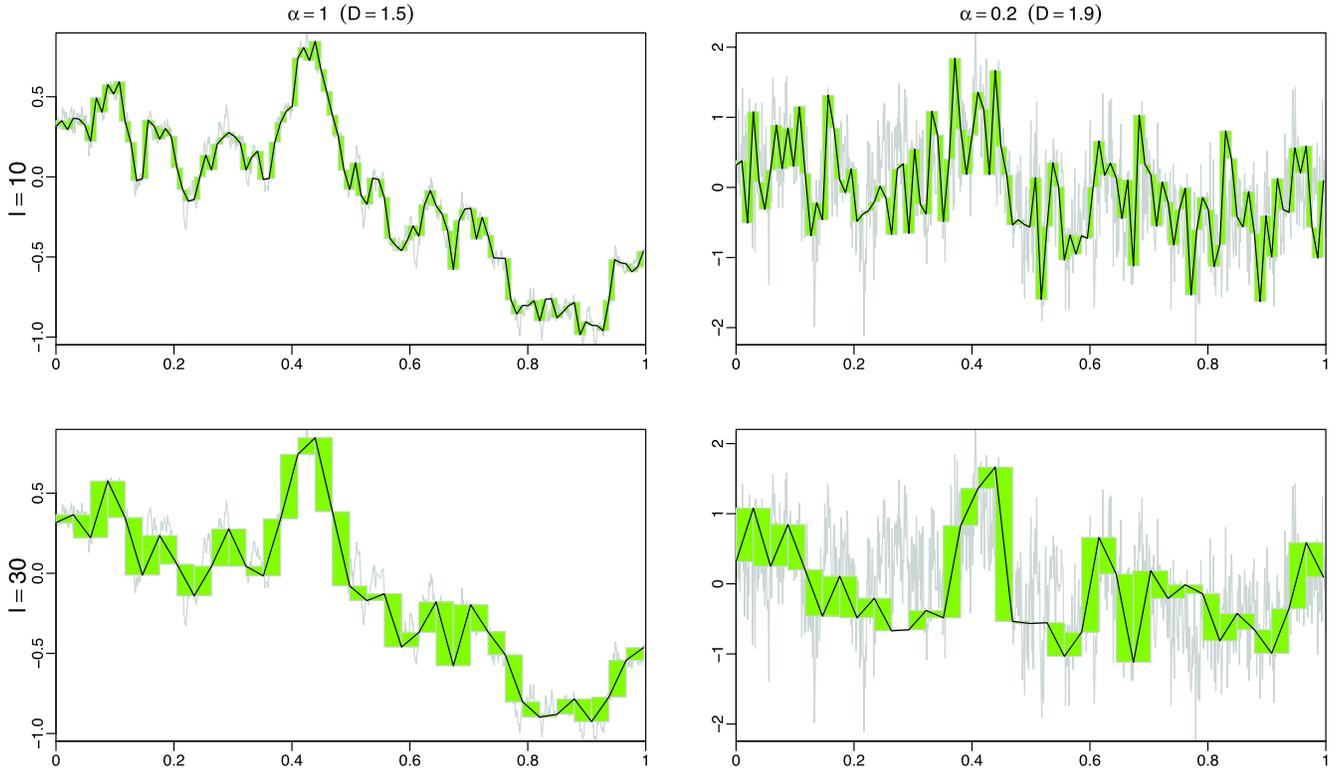}

\caption{Illustration of the Hall--Wood algorithm for the datasets in
the lower row of Figure~\protect\ref{fig1d}. The quantity $\widehat{A}(l/n)$ is
computed as the sum of the colored areas, where $n = 1{,}024$, $l =
10$ (upper row) and $l = 30$ (lower row).}
\label{fighw-howto}
\end{figure*}

\subsection{Hall--Wood Estimator} \label{sechall-wood}

Hall and Wood (\citeyear{hallwood93}) introduced a version of the box-count estimator that
operates at the smallest observed scales and avoids the need for rules
of thumb in its implementation.

To motivate their proposal, let $A(\varepsilon)$ denote the total area of
the boxes at scale $\varepsilon$ that intersect with the linearly
interpolated data graph (\ref{eqtsdata}). There are~$N(\varepsilon)$
such boxes, and so $A(\varepsilon) \propto N(\varepsilon) \hspace
{0.3mm}\varepsilon^2$,
which leads us to a reformulation of definition (\ref{eqbcfd}),
namely,
%
\begin{equation} \label{eqbcfda}
D_{\rm BC} = 2 - \lim_{\varepsilon\to0} \frac{\log A(\varepsilon)}{\log
(\varepsilon)}.
\end{equation}
At scale $\varepsilon_l = l/n$, where $l = 1, 2, \ldots,$ an estimator of
$A(l/n)$ is
%
\begin{equation} \label{eqAhat}
\widehat{A}(l/n) = \frac{l}{n}   \sum_{i=1}^{\lfloor n/ \hspace
{0.3mm}l
\rfloor}
\bigl| X_{i  l/n} - X_{(i-1)  l/n}
\bigr|  ,
\end{equation}
where $\lfloor n/  l  \rfloor$ denotes the
integer part of
$n/  l$. Figure~\ref{fighw-howto} suggests a
natural interpretation of this quantity, in that it approximates
$A(l/n)$, with all features at scales less than $l/n$ being ignored.
For an alternative, and potentially preferable, interpretation in
terms of power variations, see Section \ref{secvariation}.

\begin{figure*}[t]

\includegraphics{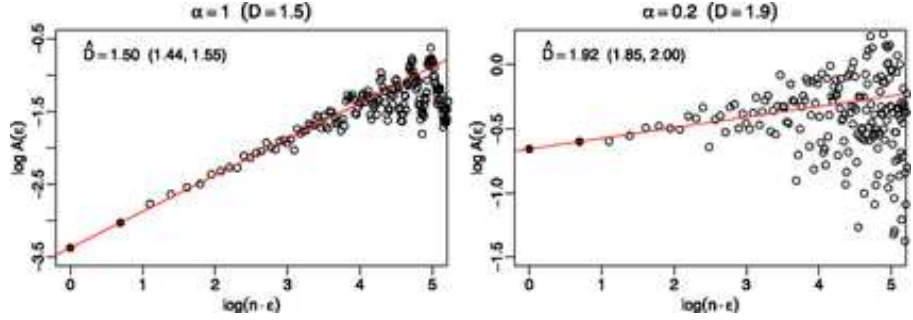}

\caption{Log--log regression for the Hall--Wood estimator (\protect\ref{eqHW})
and the datasets in the lower row of Figure~\protect\ref{fig1d}. Only the
two points marked with filled circles are used when fitting the
regression line.} \label{fighw}
\vspace*{3pt}
\end{figure*}

%
\begin{figure*}[b]
\vspace*{3pt}
\includegraphics{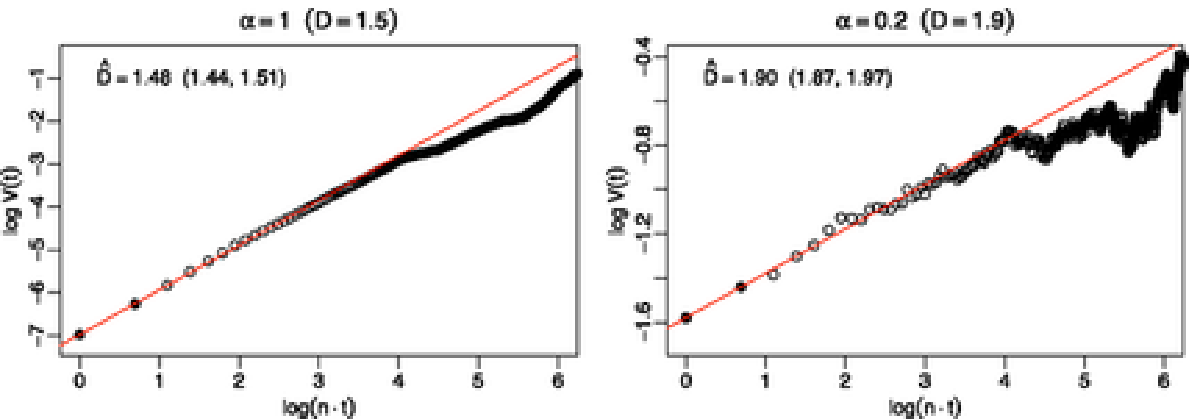}

\caption{Log-log regression for the variogram estimator (\protect\ref{eqVAR})
and the datasets in the lower row of Figure~\protect\ref{fig1d}. Only the
two points marked with filled circles are used when fitting the
regression line.}
\label{figvariogram}
\end{figure*}

The Hall--Wood estimator with design parameter $m = 1$, as used in the
numerical experiments of Hall and Wood (\citeyear{hallwood93}), is based on an ordinary
least squares regression fit of $\log\widehat{A}(l/n)$ on $\log(l/n)$:
%
\begin{equation} \label{eqFD-HW}
\widehat{D}_{\mathrm{HW}}= 2 - \Biggl\{ \sum_{l=1}^L (s_l - \bar
{s}) \log\widehat{A}(l/n)
\Biggr\}
\Biggl\{ \sum_{l=1}^L (s_l - \bar{s})^2 \Biggr\}^{-1},
\end{equation}
where $L \geq2$, $s_l = \log(l/n)$ and $\bar{s} = \frac{1}{L} \sum
_{l=1}^L s_l$.
Hall and Wood (\citeyear{hallwood93}) recommended
the use of $L = 2$ to minimize bias, which is unsurprising, in view of
the limit in (\ref{eqbcfda}) being taken as $\varepsilon\to0$. Using
$L = 2$ yields our standard implementation of the Hall--Wood estimator,
namely,
%
\begin{equation} \label{eqHW}
\widehat{D}_{\mathrm{HW}}= 2 - \frac{\log\widehat{A}(2/n) - \log
\widehat{A}(1/n)}{\log2}.
\end{equation}
Figure~\ref{fighw} shows the corresponding log--log plots and
regression fits in our illustrating example.

%

%

\subsection{Variogram Estimator} \label{secvariogram}

Owing to its intuitive appeal and ease of implementation, the
variogram estimator has been very popular. A prominent early\vadjust{\goodbreak}
application is that of Burrough (\citeyear{burrough81}). The first asymptotic study
under the infill scenario was published in the physics literature
(Jakeman and Jordan, \citeyear{JJ90}), followed by key contributions of Constantine and Hall (\citeyear{consthall94}),
Kent and Wood (\citeyear{kentwood97}),  Davies and Hall (\citeyear{DH99}) and Chan and Wood (\citeyear{chanwood04}) in statistical
journals.

Recall that the variogram or structure function~$\gamma(t)$ of a
stochastic process $\{ X_t \dvtx t \in\mathbb{R}\}$ with stationary
increments is defined in (\ref{eqdefvariogram}) as one-half times
the expectation of the square of an increment at lag $t$. The
classical method of moments estimator for $\gamma(t)$ at lag $t = l/n$
from time series or line transect data~(\ref{eqtsdata}) is
%
\begin{equation} \label{eqMoM-variogram}\quad
\widehat{V}_2(l/n) = \frac{1}{2  (n-l)}   \sum
_{i=l}^n \bigl(
X_{i/n} - X_{(i-l)/n} \bigr)^2.
\end{equation}
In view of the relationship (\ref{eqDalpha}) between the fractal
index, defined in (\ref{eqvariogram}), and the fractal dimension,
$D$, a~regression fit of $\log\widehat{V}(t)$ on $\log t$ yields the
variogram estimator,
\begin{eqnarray} \label{eqFD-variogram}
\widehat{D}_{{\mathrm V}; 2}&=& 2 - \frac{1}{2}
\Biggl\{ \sum_{l=1}^L (s_l - \bar{s}) \log\widehat{V}_2(l/n)
\Biggr\}\nonumber
\\[-8pt]\\[-8pt]
&&{}\hspace*{15pt}\cdot
\Biggl\{ \sum_{l=1}^L (s_l - \bar{s})^2 \Biggr\}^{-1},\nonumber
\end{eqnarray}
where $L \geq2$, $s_l = \log(l/n)$ and $\bar{s}$ is the mean of
$s_1, \ldots, s_L$. Figure~\ref{figvariogram} illustrates the
log--log regression for the datasets in the lower row of
Figure~\ref{fig1d}. In addition to the corresponding point estimate,
we provide a 90\% central interval estimate, using the parametric
bootstrap method as proposed by Davies and Hall (\citeyear{DH99}).


%

Constantine and Hall (\citeyear{consthall94}) argued that the bias of the variogram estimator
increases with the cut-off $L$ in the log--log regression, and
Davies and Hall (\citeyear{DH99}) showed\vadjust{\goodbreak} numerically that the mean squared error (MSE) of the
estimator for a Gaussian process with powered exponential covariance
is minimized when $L = 2$. \citet{zhustein02} argued similarly in a
spatial setting. These results are unsurprising and have intuitive
support from the fact that the behavior of the theoretical variogram
in an infinitesimal neighborhood of zero determines the fractal
dimension of the Gaussian sample paths. We thus choose $L = 2$ in our
implementation, resulting in the estimate
%
\begin{equation} \label{eqVAR}\quad
\widehat{D}_{{\mathrm V}; 2}= 2 - \frac{1}{2}   \frac{\log\widehat
{V}_2(2/n) - \log
\widehat{V}_2(1/n)}{\log2}.
\end{equation}
As Kent and Wood (\citeyear{kentwood97}) suggested, generalized least squares rather than
ordinary least squares regression could be employed, though the methods
coincide when $L = 2$.

It is well known that the method of moments estimator
(\ref{eqMoM-variogram}) is nonrobust. It is therefore tempting to
replace it by the highly robust variogram estimator proposed by
Genton (\citeyear{genton98}), which is based on the robust estimator of scale of
Rousseeuw and Croux (\citeyear{rousscroux93}). We implemented an estimator of the fractal
dimension that uses (\ref{eqFD-variogram}) with $L = 2$, but with the
method of moments estimate (\ref{eqMoM-variogram}) replaced by
Genton's highly robust variogram estimator. In a simulation setting,
this estimator works well. However, it breaks down frequently when
applied to real data, where it yields very limited, discrete sets of
possible values for the estimate only. Upon further investigation,
this stems from the ubiquitous discreteness of real-world data, under
which the Rousseeuw and Croux (\citeyear{rousscroux93}) estimator can fail, in ways just
described. While discreteness is a general issue when estimating
fractal dimension, the problem is exacerbated by the use of this
estimator. In this light, the next section investigates another
approach to more robust estimators of fractal dimension.

\subsection{Variation Estimators} \label{secvariation}

We now discuss a generalization of the variogram estimator that is
based on the variogram of order $p$ of a stochastic process with
stationary increments, namely,
%
\begin{equation} \label{eqdefvariation}
\gamma_{ p}(t) = \tfrac{1}{2}   {\mathbb{E}}|
X_u - X_{t+u}
|^{ p}  .
\end{equation}
When $p = 2$, we recover the variogram (\ref{eqdefvariogram}), when
$p = 1$ the madogram, and when $p = 1/2$ the rodogram (Bruno and Raspa, \citeyear{BR89}; Emery, \citeyear{Emery05}; Bez and Ber\-trand \citeyear{BB10}).
Standard arguments show that a Gaussian process
with a variogram of the form (\ref{eqvariogram}) admits analogous
expansions of the variogram of order $p$, in that\looseness=-1
%
\begin{eqnarray} \label{eqexpansion}
\ \gamma_{ p}(t) = |c_{ p}
 t|^{\alpha p/2}
+ \mathcal{O}\bigl(|t|^{(\alpha+\beta)  p/2}\bigr) \quad\mbox{as }
 t
\to0,\hspace*{-25pt}
\end{eqnarray}\looseness=0
with fixed values of the fractal index, $\alpha\in(0,2]$,\break $\beta>
0$, and a constant $c_{ p} > 0$ that satisfies
\[
c_{ p} = \biggl( \frac{2^{ p-1}}{\sqrt
{\pi}}   \Gamma
\biggl( \frac{p+1}{2} \biggr)
\biggr)^{2/(\alpha p)} c_2.
\]
The fractal index, $\alpha$, of the Gaussian process and the Hausdorff
dimension, $D$, of its sample paths then admit the linear relationship
(\ref{eqDalpha}).

A natural generalization of the method of moments variogram estimator
(\ref{eqMoM-variogram}) for time series or line transect data of the
form (\ref{eqtsdata}) is the power variation of order $p$, namely,
%
\begin{equation} \label{eqMoM-variation}\qquad
\widehat{V}_p(l/n) = \frac{1}{2  (n-l)}
\sum_{i=l}^n \bigl| X_{i/n} - X_{(i-l)/n} \bigr|^{ p}.
\end{equation}
We then define the variation estimator of order $p$ for the fractal
dimension as
\begin{eqnarray} \label{eqFD-variation}
\widehat{D}_{{\mathrm V}; p}&=& 2 - \frac{1}{p}
\Biggl\{ \sum_{l=1}^L (s_l - \bar{s}) \log\widehat{V}_p(l/n)
\Biggr\}\nonumber
\\[-8pt]\\[-8pt]
&&{}\hspace*{15pt}\cdot\Biggl\{ \sum_{l=1}^L (s_l - \bar{s})^2 \Biggr\}^{-1},\nonumber
\end{eqnarray}
where $L \geq2$, $s_l = \log(l/n)$ and $\bar{s}$ is the mean of
$s_l, \ldots, s_L$. This definition nests the variogram, madogram and
rodogram estimators, which arise when $p = 2$, $1$ and $1/2$,
respectively. The general case, $p > 0$, has been studied by
 Coeurjolly (\citeyear{coeurjolly01}, \citeyear{coeurjolly08}).

\begin{figure*}[t]

\includegraphics{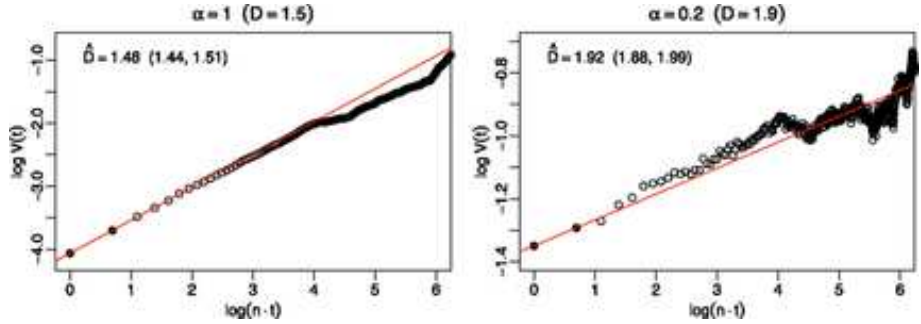}

\caption{Log--log regression for the madogram estimator [variation
estimator (\protect\ref{eqV})
with power index $p = 1$] and the datasets in the lower row of
Figure~\protect\ref{fig1d}. Only the two points marked with filled circles
are used when fitting the regression line.}
\label{figvariation}
\vspace*{12pt}
\end{figure*}

For the same reasons as before, and supported by
simulation experiments, we let $L = 2$ in our implementation, so that
%
\begin{equation} \label{eqV}
\quad \widehat{D}_{{\mathrm V}; p}= 2 - \frac{1}{p}
\frac{\log\widehat{V}_p(2/n) - \log\widehat{V}_p(1/n)}{\log2}.
\end{equation}
As an illustration, Figure \ref{figvariation} shows the log--log
regression fit for the variation estimator of order $p = 1$ and our
example data. For instances of the use of the madogram estimator in
the applied literature, see~Weissel, Pratson and Malinverno (\citeyear{weisseletal94}) and
Zaiser et~al. (\citeyear{zaiser04}).

%
%

A natural question then is for the choice of the power index $p > 0$.
With the estimator depending on the relationship (\ref{eqDalpha})
between the fractal index, $\alpha$, in the expansion
(\ref{eqexpansion}) and the Hausdorff dimension,~$D$, of the sample
path, it is critically important to assess its validity when the
assumption of Gaussianity is violated. In the standard case in which
$p = 2$, Hall and Roy (\citeyear{HR94}) showed that, while the relationship~(\ref{eqDalpha}) extends to some non-Gaussian processes, it fails
easily. For example, it applies to marginally power-transformed
Gaussian fields if and only if the transformation power exceeds $1/2$.
Other counterexamples can be found in the work of Bruno and Raspa (\citeyear{BR89}) and
Scheuerer (\citeyear{Scheuerer10}).

Bruno and Raspa (\citeyear{BR89}) and Bez and Ber\-trand (\citeyear{BB10})
applied the Lipschitz--H\"older heuristics
of\break Mandelbrot [(\citeyear{Mandelbrot77}), page~304] to argue that the
relationship (\ref{eqDalpha}) is universal when $p = 1$. While we
agree that the relationship is particularly robust and inclusive when
$p = 1$, the Lipschitz--H\"older heuristics, which connects the
Lipschitz exponent of a sample path to its Hausdorff dimension, is
tied to continuity. Thus, it can fail if the sample paths are
sufficiently irregular. For instance, the sample paths of a binary
stochastic process, which attains the values 0 and 1 only, have
Hausdorff dimension $D = 1$. As the corresponding variogram
(\ref{eqdefvariation}) is independent of its order, we may refer to
the common version as $\gamma$. If the expected number of sample path
jumps per unit time is finite, then $\gamma$ grows linearly at the
coordinate origin (Masry, \citeyear{Masry72}). In this case, the
relationship~(\ref{eqDalpha}) holds if, and only if, the common
variogram,~$\gamma$, is~un\-derstood to be of order $p = 1$. However, there are
binary processes whose variogram behaves like $\gamma(t) = \mathcal{
O}(|t|^\gamma)$ as $t \to0$, where $0 < \gamma< 1$, and then the
relationship fails. Notwithstanding these examples, the argument of
Bruno and Raspa (\citeyear{BR89}) and Bez and Ber\-trand (\citeyear{BB10}) is persuasive, and we maintain that the
relationship (\ref{eqDalpha}) is particularly inclusive when $p = 1$.
A natural conjecture is that if $p = 1$, then the relationship is valid
if the process is ergodic (in a~suitable sense) and the expected
number of sample path discontinuities per unit time is
finite. Furthermore, it is worth noting that there are processes for
which the madogram exists and the foregoing is satisfied, while second
moments do not exist (Ehm, \citeyear{Ehm81}).

The following interesting connection between the Hall--Wood estimator
and the madogram estimator also suggests a special role of the power
index $p = 1$. For $l$ a positive integer and $j = 0, 1, \ldots, l
-1$, let
\[
\widehat{A}^{  (  j)}(l/n)
= \frac{l}{n}   \sum_{i=1}^{\lfloor(n-j)/  l
\rfloor}
\bigl| X_{(i  l+j)/n} - X_{(i  l+j-l)/n}
\bigr|  .\vadjust{\goodbreak}
\]
Then $\widehat{A}(l/n) = \widehat{A}^{  (0)}(l/n)$ and
\[
\widehat{V}_1(l/n)
= \frac{1}{2}   \frac{n}{n-1}   \frac{1}{l}
\sum_{j=0}^{l-1} \widehat{A}^{  (  j)}(l/n)
\]
is, up to inessential constants, the mean of $l$ distinct copies of
$\widehat{A}(l/n)$. A comparison of the general forms~(\ref{eqFD-HW}) and (\ref{eqFD-variation}), or the standard forms
(\ref{eqHW}) and~(\ref{eqV}), of the Hall--Wood estimator with the
madogram estimator suggests
that the latter is a statistically more efficient version
of the Hall--Wood estimator. A similar, more tedious calculation can
be used to argue heuristically that the box-count estimator has a
bias, typically leading to lower estimates of the fractal dimension
than the Hall--Wood and variation estimators. For a confirmation in
simulation studies, see Section \ref{secsimulation1d}.

%
\begin{figure*}[t]

\includegraphics{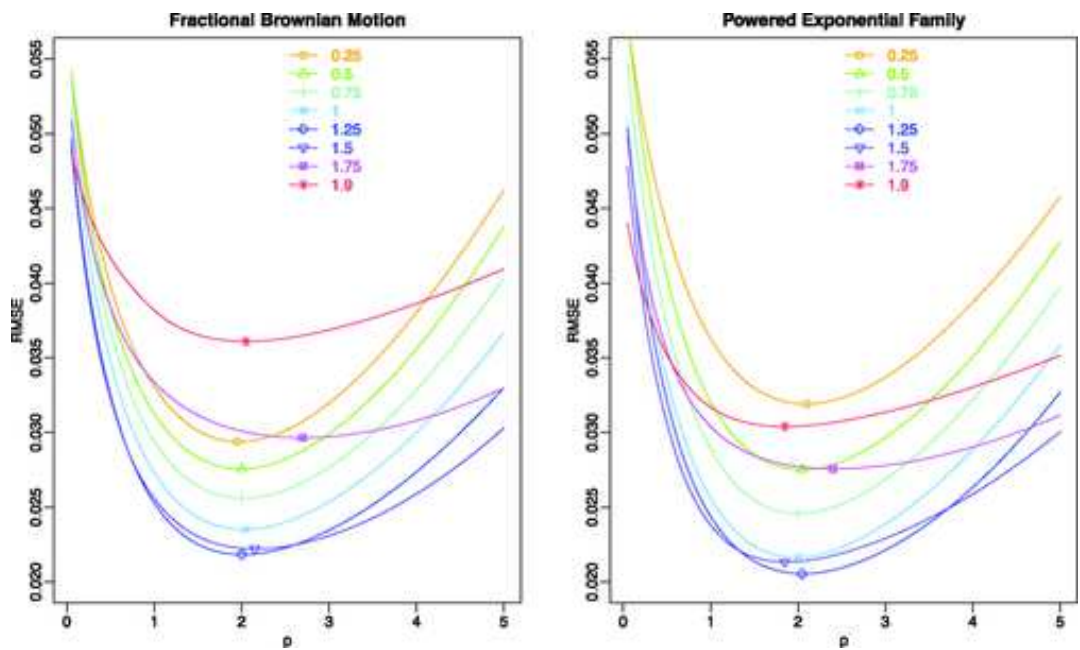}
\caption{Root mean squared error (RMSE) of the variation estimator
(\protect\ref{eqV}) as a function of the power exponent, $p$, for Gaussian
fractional Brownian motion (left) and Gaussian processes with
powered exponential covariance (right). The scale parameter used is
$c = 1$, and each RMSE is computed from 1,000 Monte Carlo
replicates of Gaussian sample paths under the sampling scheme
(\protect\ref{eqtsdata}), where $n = 1{,}024$. The curves correspond to
fixed values of the fractal index, $\alpha$, and have their minima
marked.}
\label{figHE-p-alpha}\vspace*{12pt}
\end{figure*}

\begin{figure*}[t]

\includegraphics{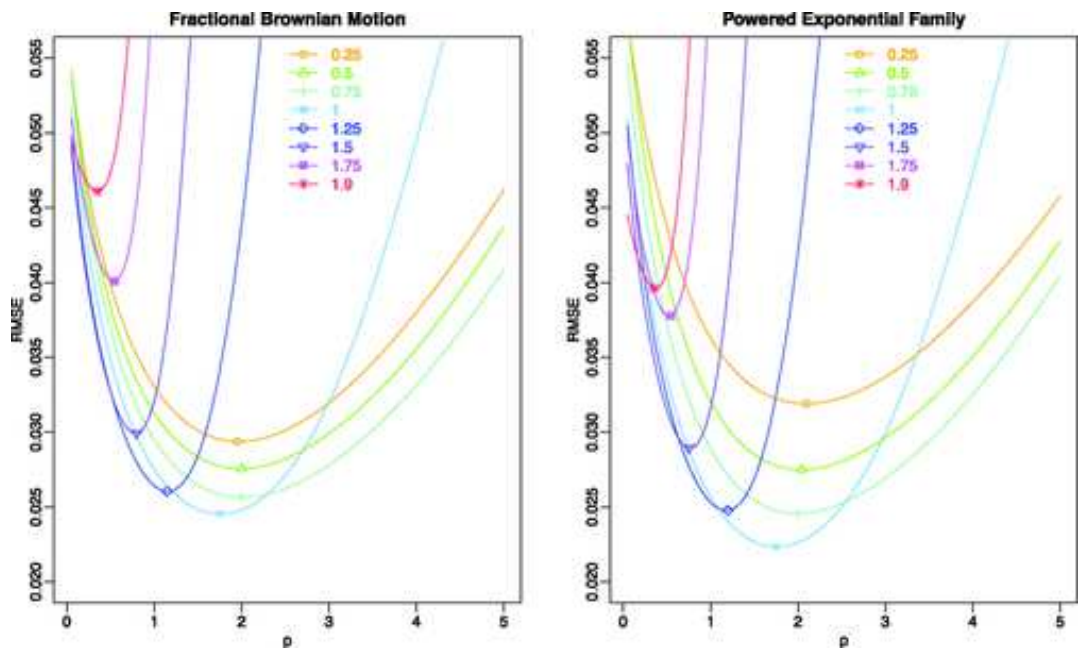}

\caption{Same as Figure \protect\ref{figHE-p-alpha}, except that in each
sample path a randomly placed observation is contaminated by an
additive Gaussian outlier with standard deviation $0.1$.}
\label{figHE-outl-p-alpha}\vspace*{24pt}
\end{figure*}
%

Here we restrict attention to a small initial study that assesses the
efficiency and outlier resistance of variation estimators. Figures
\ref{figHE-p-alpha} and \ref{figHE-outl-p-alpha} show the root mean
squared error (RMSE) of the variation estimator (\ref{eqV}) from
Gaussian sample paths in dependence on the power index, $p$. The
curves are computed from 1,000 independent realizations with sample
size $n = 1{,}024$, correspond to fixed values of the fractal index,
$\alpha$, and have their minima marked. Figure \ref{figHE-p-alpha}
concerns the ideal Gaussian model, where the estimator performs best
for power indices of about $p = 2$, corresponding to the variogram
estimator, similarly to the observations of
Coeurjolly [(\citeyear{coeurjolly01}), page~1417]. Figure \ref{figHE-outl-p-alpha} shows
that the RMSE can deteriorate considerably under a single additive
outlier, with the effect being stronger for smoother sample paths,
that is, higher values of the fractal index. The smaller the power
index, the more outlier resistant the variation estimator.

A possible variant of the variation estimator uses $p$-moments of
higher increments, as proposed by Kent and Wood (\citeyear{kentwood97}) and
Istas and Lang (\citeyear{istaslang97}) in the case $p = 2$. For example, one could turn
to second differences, rather than first differences, and base a~log--log regression on
\begin{eqnarray} \label{eqMoM-variation-1}
\quad \widehat{V}_p^{(2)}(l/n) &=& \frac{1}{2  (n-2l)}\nonumber
\\[-8pt]\\[-8pt]
&&{}\cdot  \sum
_{i=l}^{n-l}
\bigl| X_{(i+l)/n} - 2 X_{i/n} + X_{(i-l)/n} \bigr|^{ p},\nonumber
\end{eqnarray}
rather than the standard variation (\ref{eqMoM-variation}). Also, if
more than two points are used in the regression, the generalized least
squares method could be used in lieu of the ordinary least squares
technique. However, there is no clear advantage in doing so in
applied settings, in which the corresponding covariance structure is
unknown and needs to be estimated as well.

\subsection{Spectral and Wavelet Estimators} \label{secperiodogram}

\begin{figure*}
\includegraphics{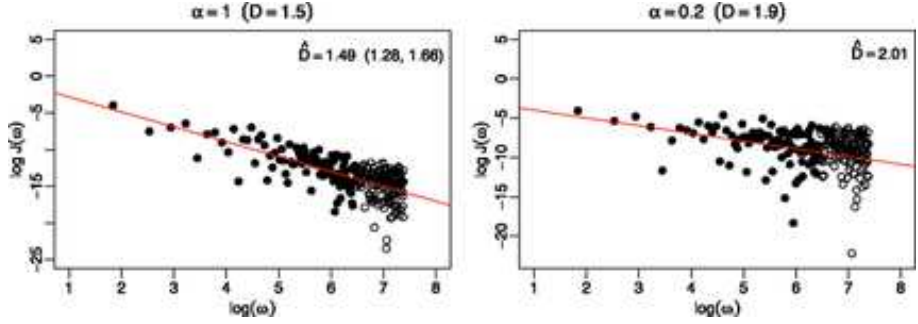}

\caption{Log-log regression for the semi-periodogram estimator and the
datasets in the lower row of Figure~\protect\ref{fig1d}. Only the points
marked with filled circles are used when fitting the regression
line.}
\label{figperiodogram}
\end{figure*}

We now consider the semi-periodogram estimator of Chan, Hall and Poskitt (\citeyear{chanhall95}),
which operates in the frequency domain and is closely related to the
spectral estimator of Dubuc et~al. (\citeyear{dubuc89a}). The basis for this estimator
is the well-known fact that the spectral density function for a
stationary stochastic process with a second-order variogram of the
form~(\ref{eqvariogram}) decays like $|\omega|^{-\alpha-1}$ as
frequency $|\omega| \to\infty$ (Stein, \citeyear{stein99}). For a stationary
Gaussian process $\{ X_t \dvtx t \in[0,1] \}$, Chan, Hall and Poskitt
(\citeyear{chanhall95}) defined
\[
B(\omega) = 2 \int_{0}^1 X_t \cos(\omega[2t-1]) \,  {\rm d}t
\]\looseness=0
and called
\[
J(\omega) = B(\omega)^2
\]
the semi-periodogram. Under weak regularity conditions, the expected
value of the semi-periodogram decays in the same way as the spectral
density function. Suppose now that we have $n_s = 2m+1$ observations
$X_t$ at times $t = i/(2m)$, where $i = 0, 1, \ldots, 2m.$
In this setting, Chan, Hall and Poskitt (\citeyear{chanhall95}) approximated $B(\omega)$ by
\[
\widehat B(\omega) = \frac{1}{m}
\Biggl[ \frac{X_{0} + X_1}{2}
+ \sum_{i=1}^{2m-1} X_{i/(2m)} \cos\biggl( \omega
\frac{i-m}{m}
\biggr) \Biggr]
\]
and the semi-periodogram $J(\omega)$ by
\[
\widehat{J}(\omega) = \widehat B(\omega)^2.
\]
The semi-periodogram estimator of the fractal dimension $D$ is
\begin{eqnarray} \label{eqfdperiodogram}
\widehat{D}_{{\mathrm P}}&=& \frac{5}{2} + \frac{1}{2}
\Biggl\{ \sum_{l=1}^L (s_l - \bar{s}) \log\widehat{J}(\omega_l)
\Biggr\}\nonumber
\\[-8pt]\\[-8pt]
&&{}\hspace*{16pt}\cdot
\Biggl\{ \sum_{l=1}^L (s_l - \bar{s})^2 \Biggr\}^{-1},\nonumber
\end{eqnarray}
where $\omega_l = 2 \pi l$, $s_l = \log\omega_l$ and $\bar{s}$ is the
mean of~$s_1,\allowbreak \ldots, s_L$.
The highest unaliased frequency (i.e., the~Ny\-quist
frequency) is $\pi m$, which is reflected by the fact that $\widehat
B(\pi m + \delta) = \widehat B(\pi m - \delta)$
for any $\delta$.
This suggests setting $L=\lfloor m/2\rfloor$,
but Chan, Hall and Poskitt (\citeyear{chanhall95}) recommended using $L = \lfloor\min\{m/2,
n_s^{2/3}\}\rfloor$, which
is less than $\lfloor m/2\rfloor$ for $m\ge34$. As $m$ grows, this
rule thus has the
effect of eliminating $\widehat{J}(\omega)$ at high frequencies in
forming $\widehat{D}_{{\mathrm P}}$, which at first seems
counterintuitive, given that the
underlying scaling law applies as $\omega$ increases. However, as
$\omega$ approaches the Nyquist frequency, aliasing causes the
expectation of $\widehat{J}(\omega)$ to deviate markedly from the
decay rate of $\omega^{-\alpha-1}$, leading to the need to eliminate
high-frequency terms when forming $\widehat{D}_{{\mathrm P}}$. Figure
\ref{figperiodogram}
shows an example of the log--log regression for the semi-periodogram
estimator for datasets from Figure~\ref{fig1d}, where $n_s = 1{,}025$
and hence $L = 101 < m/2 = 256$.

%

\begin{figure*}[b]

\includegraphics{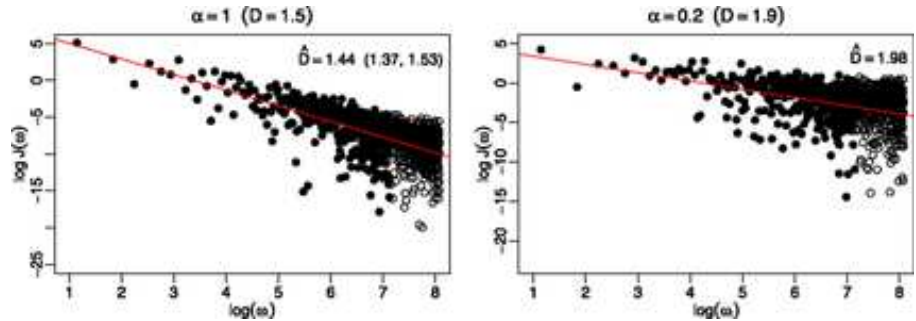}

\caption{Log--log regression for the DCT-II estimator and the datasets
in the lower row of Figure~\protect\ref{fig1d}. Only the points marked with
filled circles are used when fitting the regression line.}
\label{figdctII}
\end{figure*}

%

The definition of $\widehat B(\omega)$ is similar in spirit to the
so-called type-II discrete cosine transform (DCT-II); see Ahmed, Natarajan and Rao
(\citeyear{ahmed74})
and Strang (\citeyear{strang99}) for background. Davies (\citeyear{davies01})
noted that this transform has some attractive properties when used as
a basis for spectral analysis, so it is of interest to explore the
DCT-II as a substitute for the semi-periodogram in estimating fractal
dimension. Given time series data of the form (\ref{eqtsdata}),
taking the definition of the DCT-II given by Gonzalez and Woods
(\citeyear{gonzalezwoods07})
and adjusting it for our convention for the samples $X_{i/(2m)}$ yields
\[
\widetilde B(\omega) = \biggl( \frac{2}{2m+1} \biggr)^{1/2}
\sum_{i=0}^{2m} X_{i/(2m)} \cos\biggl(\omega \frac{2i+1}{4m}
\biggr)
\]
and $\widetilde{J}(\omega) = \widetilde B(\omega)^2$.
Here the Nyquist frequency is $2\pi m$,
as can be seen by noting that
$\widetilde J(2\pi m + \delta) = \widetilde J(2\pi m - \delta)$ for
any $\delta$.
If we now let
$\omega_l = 2\pi lm/(2m+1)$ with $s_l = \log\omega_l$ and $\bar{s}$
being the mean of the $s_l$'s as before, the log--log regression\vadjust{\goodbreak}
estimator (\ref{eqfdperiodogram}) can be applied with
$\widehat{J}(\omega_l)$ replaced by $\widetilde{J}(\omega_l)$ and
$L =
\lfloor\min  \{ 2m, 4n_s^{2/3} \}\rfloor$.
The DCT-II estimator uses approximately four times more
points in the log--log regression than does the semi-periodogram estimator.
For example, in Figure~\ref{figdctII} we have $L = 406$, whereas we had
$L =101$ in Figure~\ref{figperiodogram}.

\begin{figure*}

\includegraphics{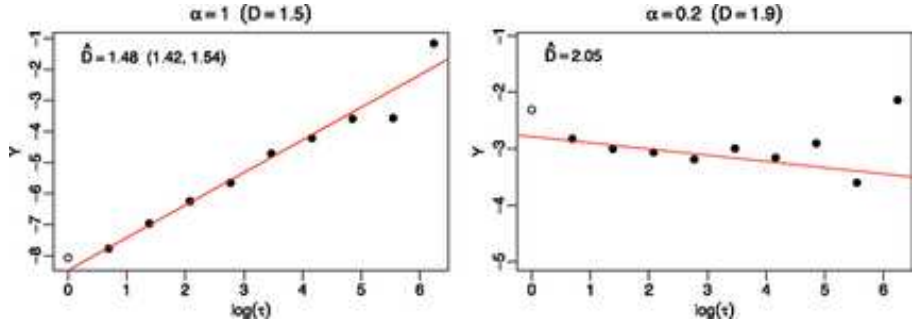}

\caption{Log--log regression for the wavelet estimator and the datasets
in the lower row of Figure~\protect\ref{fig1d}. Only the points
marked with filled circles are used when fitting the regression
line.} \label{figwavelet}
\end{figure*}

The semi-periodogram estimator also serves as motivation for a similar
wavelet estimator, which is an adaptation of a weighted least squares
estimator for the long memory parameter of a fractionally differenced
process (Percival and Walden, \citeyear{percivalwalden00}, Section~9.5).
Given a time series of
length $n_s$, we compute its maximal overlap discrete wavelet
transform (MODWT) out to level $J_0 = \lfloor\log_2(n_s) \rfloor$
using reflection boundary conditions; this can be done using the
function {\tt modwt} in the \texttt{R} package {\tt wavelets}~(Aldrich, \citeyear{wavelets10}). This
MODWT yields $J_0$ vectors of wavelet coefficients $\widetilde{\bf
W}_j$, $j = 1, \ldots, J_0$, each of which contains~$2n_s$
coefficients. The coefficients in the $j$th vector are associated
with the scale $\tau_j = 2^{j-1}$. The average of these coefficients
squared, that is, $\| \widetilde{\bf W}_j \|^2/2n_s$, provides an
estimator of the wavelet variance $\nu^2(\tau_j)$. This variance
varies approximately as $\tau_j^\alpha$ for large $\tau_j$
[Percival and Walden, \citeyear{percivalwalden00},\vadjust{\goodbreak}
equation~(297b)],\break where~$\alpha$ is the fractal
index, from which the fractal dimension can be deduced; see
Table~\ref{tabfdmethods}. The scale $\tau_j$ corresponds to the band
of frequencies $(\pi/2^j, \pi/2^{j-1}]$.

The information that is captured by the semi-perio\-dogram at high
frequencies is thus captured in the wavelet variance at small scales
$\tau_j$. Since Chan, Hall and Poskitt (\citeyear{chanhall95}) eliminated certain high frequencies
in their semi-periodogram estimator, this suggests using just the
wavelet variances indexed by $j = J_0, \ldots, J_1$, where $J_0 = \max
  \{ 1, \lfloor\log_2(n_s)/3 - 1 \rfloor\}$. Because the variance
of the wavelet variance estimators depends upon $\tau_j$, we replace
the ordinary least squares estimator of the slope that is the basis
for equation~(\ref{eqfdperiodogram}) with a weighted least squares
estimator, say $\hat\alpha_{\rm WL}$ [Percival and Walden, \citeyear{percivalwalden00}, equation~(376c)]. The corresponding estimator of the fractal
dimension $D$ is $\widehat{D}_{\rm WL} = 2 - \frac
{1}{2}\hat
\alpha_{\rm WL}$. Figure~\ref{figwavelet} gives an
example of
the wavelet estimator of $D$. Note that the estimator of $D$ in the
right-hand plot is $\widehat{D}_{\rm WL} = 2.05$, which is
greater than the upper limit $D=2$ for the Hausdorff dimension of a curve.
Sampling variability can cause
this to happen on occasion with the other estimators also. In the
simulation experiments reported below, the wavelet estimator proved to
perform comparably to the DCT-II estimator, so we have chosen to drop
the former and report only on the latter in what follows.

\section{Performance Assessment: Time Series and Line Transect Data}
\label{secperformance1d}

We now turn to an evaluation of the various types of estimators, where
we consider the large sample behavior under infill asymptotics and
report on a~finite sample simulation study that assesses both
efficiency and robustness.

\subsection{Asymptotic Theory} \label{secasymptotic1d}

As noted, the fractal dimension refers to the properties of a graph in
a hypothetical limiting process that might exist if the scale of
measurement were to become infinitely fine. Hence, estimators of
fractal dimension are studied under infill asymptotics
(Hall and Wood, \citeyear{hallwood93}; Stein, \citeyear{stein99}), in which the number of observations grows
to infinity, whereas the underlying domain, namely the unit interval,
remains fixed. We assume that time series or line transect data of
the form (\ref{eqtsdata}) arise from a Gaussian process $\{ X_t \dvtx t
\in[0,1] \}$ with a second-order structure of the type
(\ref{eqvariogram}), where we let $n$ grow without bounds.
Typically, the literature assumes stationarity, so that the process
$\{ X_t \dvtx t \in[0,1] \}$ has a covariance function of the form
\[
\sigma(t) = \sigma(0) - |c   t|^\alpha+ \mathcal{O}(|t|^{\alpha+\beta})
\quad\mbox{as }  t \to0,
\]
where $\alpha\in(0,2)$, $\beta\geq0$ and $c > 0$. The behavior of
the bias, variance and mean squared error (MSE) of the estimators, and
the corresponding types of limit distributions, then depend on the
fractal index~$\alpha$ and on $\beta$. Typically, the corresponding
asymptotic results carry over to Gaussian processes with stationary
increments and a variogram or structure function of the form
(\ref{eqvariogram}).\vadjust{\goodbreak}

For any Hall--Wood or variogram estimator $\widehat{D}$ of the form
(\ref{eqFD-HW}) or (\ref{eqFD-variogram}) with a fixed value of the
design parameter $L$, the key results of Hall and Wood (\citeyear{hallwood93}) and
Constantine and Hall (\citeyear{consthall94}) are that
\begin{eqnarray} \label{eqasy1d}
\quad\hspace*{8pt} &&\mbox{MSE}(\widehat{D})\nonumber
\\[-8pt]\\[-8pt]
&&\quad =
\cases{
\mathcal{O}(n^{-1}) + \mathcal{O}(n^{-2\beta}), & \mbox{if }$  0 < \alpha<
\frac{3}{2}$,
\cr
\mathcal{O}(n^{-1} \log n) + \mathcal{O}(n^{-2\beta}), & \mbox{if }$  \alpha
= \frac{3}{2}$,
\cr
\mathcal{O}(n^{2\alpha-4}) + \mathcal{O}(n^{-2\beta}), & \mbox{if }$  \frac
{3}{2} < \alpha< 2$,
}\nonumber\hspace*{-5pt}
\end{eqnarray}
where in each case the first term corresponds to the variance, and the
second term to the squared bias. If $\alpha\leq3/2$, then
$\widehat{D}$ has a normal limit; if $\alpha> 3/2$, the limit is a
Rosenblatt distribution as described by Taqqu (\citeyear{Taqqu75}). In a recent
far-reaching paper, Coeurjolly (\citeyear{coeurjolly08}) showed that in the Gaussian
case and for the variation estimator (\ref{eqFD-variation}) with
general power index $p > 0$, the asymptotic behavior is still
described by (\ref{eqasy1d}). Furthermore, the convergence rates
are retained if the arithmetic mean in the definition of the power
variation (\ref{eqMoM-variation}) is replaced by a trimmed mean, or
by a convex combination of sample quantiles. While some of these
results carry over to certain specific non-Gaussian processes
(Chan and Wood, \citeyear{chanwood04}; Achard and Coeurjolly, \citeyear{achard10}), the limiting distribution theory is
considerably richer then, and a~general non-Gaussian theory remains
lacking.

\begin{figure*}[b]
\vspace*{6pt}
\includegraphics{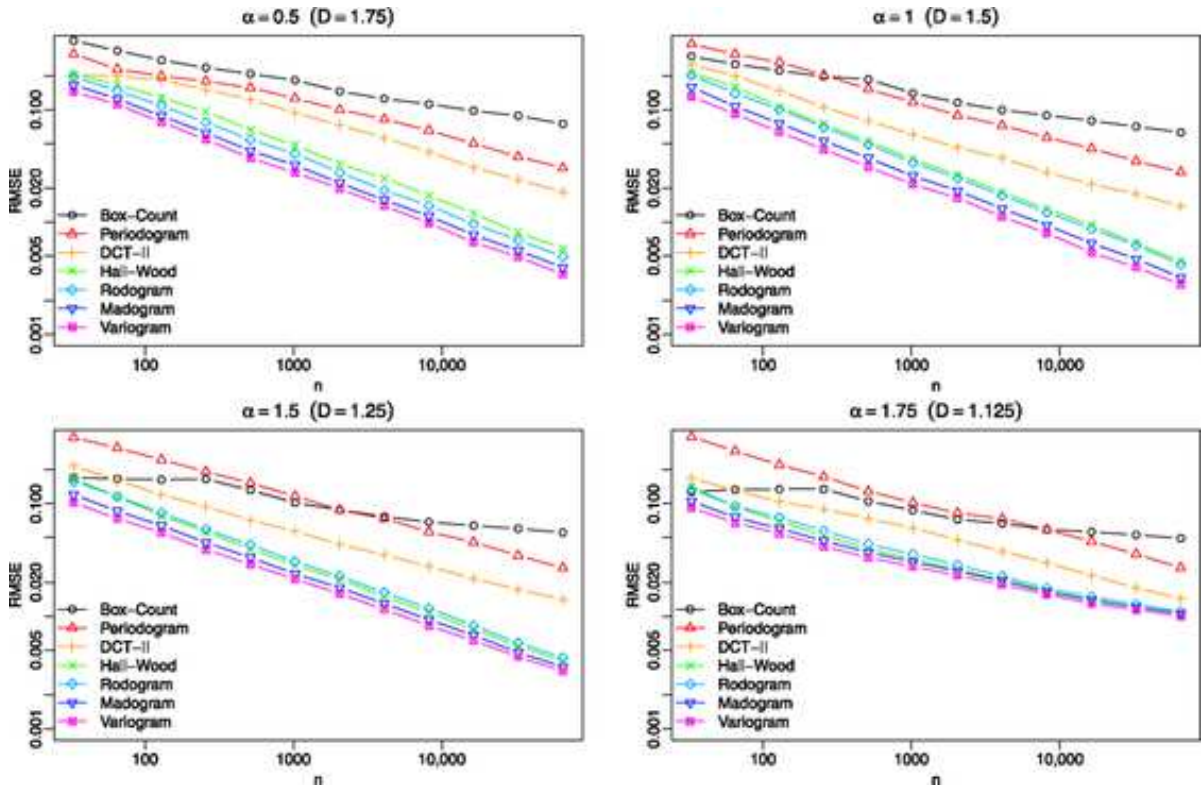}

\caption{Root mean squared error (RMSE) of estimators of fractal
dimension in dependence on the sample size, $n$, computed from
Gaussian sample paths of the form (\protect\ref{eqtsdata}) with powered
exponential covariance function, $\sigma(t) = \exp(-|t|^\alpha)$.
For each combination of $\alpha$ and $n$, the number of Monte Carlo
replicates is 1,000.} \label{fig1d-fbm-rmse}
\end{figure*}

It is interesting to observe the change in the asymptotic rate of
convergence at $\alpha= 3/2$ for all these types of estimators.
However, Kent and Wood (\citeyear{kentwood97}) showed that the variogram estimator
achieves an MSE of order
%
\begin{equation} \label{eqasybest}
\mbox{MSE}(\widehat{D}) =\mathcal{O}(n^{-1})
+ \mathcal{O}(n^{-2\beta})
\end{equation}
for all $\alpha\in(0,2)$ if the design parameter satisfies $L \geq3$ and the generalized least
squares technique, rather than the ordinary least squares method, is
used in\vadjust{\goodbreak} the log--log regression fit, and/or second differences of the
form (\ref{eqMoM-variation-1}) are used. Similarly,
Coeurjolly (\citeyear{coeurjolly08}) demonstrated that the asymptotic rate of
convergence for the variation estimator with general power index $p >
0$ can be improved if second differences or related special types of
increments are used. In finite sample simulation studies for the
variogram estimator ($ p = 2$), Kent and Wood (\citeyear{kentwood97}) did
not find a
clear-cut advantage in using generalized least squares and/or second
differences, and our own experience with variation estimators of
diverse power indices is similar. As Kent and Wood (\citeyear{kentwood97}) argued, the
likely cause is that, the closer $\alpha$ is to 2, the larger $n$ must
be before the asymptotic regime is reached. This behavior is in
marked contrast to the case of spatial lattice data, to be discussed
below.\looseness=1

Chan, Hall and Poskitt (\citeyear{chanhall95}) developed asymptotic theory for the semi-periodogram
estimator, but the MSE decays at best at rate $\mathcal{O}(n^{-1/4})$ in
their results. The asymptotic scenario for the level crossing
estimator in Feuerverger, Hall and Wood (\citeyear{feuerhallwood94}) involves a bandwidth parameter
and thus is not directly comparable.

\subsection{Simulation Study: Gaussian Processes} \label{secsimulation1d}

%

%

We now turn to a simulation study, in which we confirm and complement
the foregoing asymptotic results in a Gaussian setting. In doing so,
exact simulation is critically important (Chan and Wood, \citeyear{chanwood00}; \cite{zhustein02}), and we use the circulant embedding approach
(Dietrich and Newsam \citeyear{dietrichnewsam93}; Wood and Chan, \citeyear{woodchan94}; Stein, \citeyear{stein02}; Gneiting et~al., \citeyear{Getal06}) as implemented
in the \texttt{R} package \texttt{RandomFields} (Schlather, \citeyear{schlather01}) to
generate Gaussian sample paths, using the function \texttt{GaussRF}. The
circulant embedding technique relies on the fast Fou\-rier transform and
is both exact and fast.

\begin{figure*}

\includegraphics{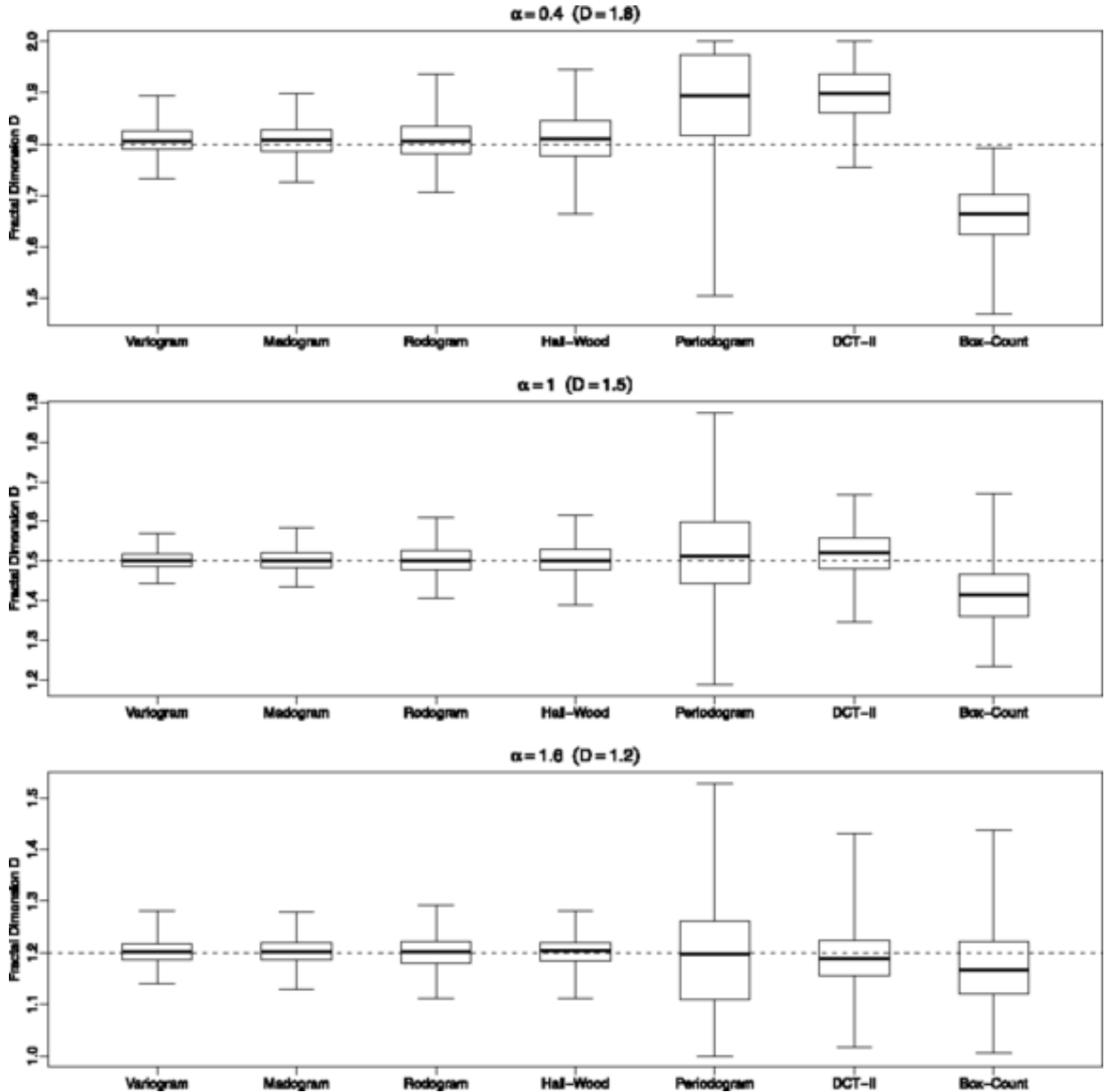}

\caption{Boxplots for estimates of the fractal dimension from Gaussian
sample paths of the form (\protect\ref{eqtsdata}), where $n = 1{,}024$, with
powered exponential covariance function, $\sigma(t) =
\exp(-|t|^\alpha)$, and the fractal index, $\alpha$, being equal to
0.4, 1.0 and 1.6, respectively. The corresponding true values of
the fractal dimension, $D$, namely 1.8, 1.5 and 1.2, are shown as
dashed lines. The number of Monte Carlo replicates is
500.} \label{figboxplots-sim}\vspace*{-3pt}
\end{figure*}

Figure \ref{fig1d-fbm-rmse} shows log--log plots for the root mean
squared error (RMSE) of the various types of estimators in their
dependence on the sample size $n$, computed from 1,000 independent
trajectories of the form (\ref{eqtsdata}) from the corresponding
stationary Gaussian process with a powered exponential covariance
function, $\sigma(t) = \exp(-|t|^\alpha)$. The graphs are
approximately linear, and their slopes show good agreement with the
asymptotic laws in (\ref{eqasy1d}). Furthermore, they confirm
the\vadjust{\goodbreak}
aforementioned observation of Kent and Wood (\citeyear{kentwood97}) that large values of
the fractal index,~$\alpha$, require large sample sizes to reach the
asymptotic regime. The variogram estimator generally shows the lowest
MSE, followed by the madogram
and rodogram, and then the Hall--Wood, DCT-II, periodogram and box-count estimators. This ranking is retained under Gaussian processes
with covariance functions from the Mat\'ern and Cauchy families, as
well as for fractional Brownian motion, for all values of the fractal
index $\alpha$ and all sufficiently large sample sizes, $n$. The use
of variation estimators based on second differences, as defined in
(\ref{eqMoM-variation-1}), typically does not yield lower RMSEs
(results not shown).

%

\begin{figure*}

\includegraphics{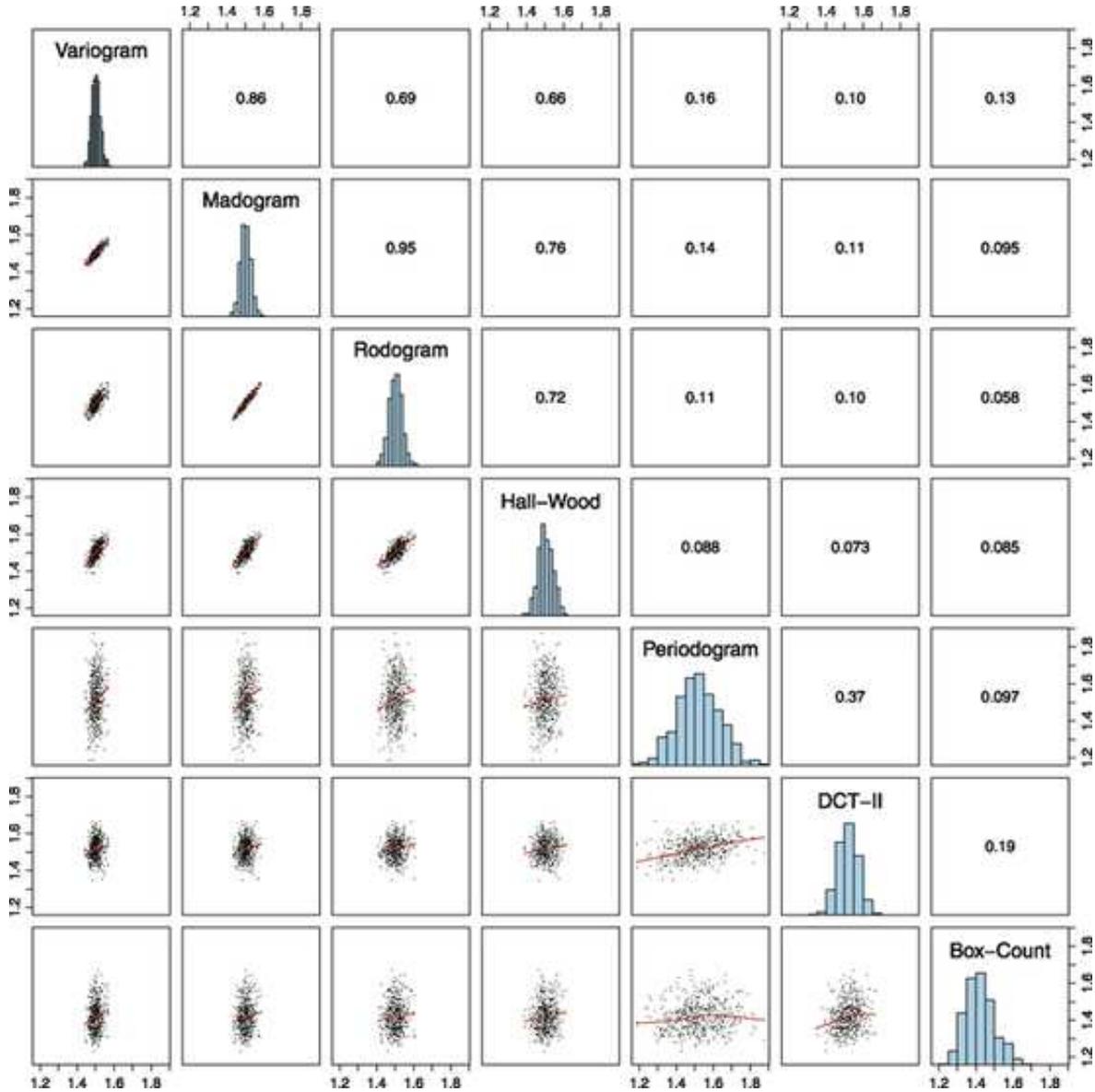}

\caption{Scatterplot matrix for estimates of the fractal dimension
from Gaussian sample paths of the form (\protect\ref{eqtsdata}) with
exponential covariance function, $\sigma(t) = \exp(-|t|)$, and
sample size $n = 1{,}024$. The true value of the fractal dimension
is $D = 1.5$. The panels along the diagonal show histograms of the
estimates, and those above the diagonal pairwise Pearson correlation
coefficients. The number of Monte Carlo replicates is 500.}
\label{figscatterplot-matrix-sim}
\end{figure*}
%

Figures \ref{figboxplots-sim} and \ref{figscatterplot-matrix-sim}
show box- and scatterplots for the same types of estimators and the
same class of Gaussian processes, where the sample size is \mbox{$n = 1{,}024$}.
Three groups of estimators can be distinguished, the first
comprising the variogram and other two variation estimators along with the
Hall--Wood estimator, the second the spectral estimators, and the third
the box-count estimator. The most efficient estimator is the
variogram estimator, closely followed by the madogram
estimator. As we have argued theoretically
\mbox{before}, the madogram estimator is a more efficient version of the
Hall--Wood estimator, in that the estimators are strongly correlated,
but the former is less \mbox{dispersed}. While the spectral estimators are
less competitive, showing substantially higher dispersion than the
variation estimators, the DCT-II estimator improves considerably on
the periodogram estimator. The box-count estimator generally shows a
bias, with the estimates being too low.

%
\begin{figure*}[t]

\includegraphics{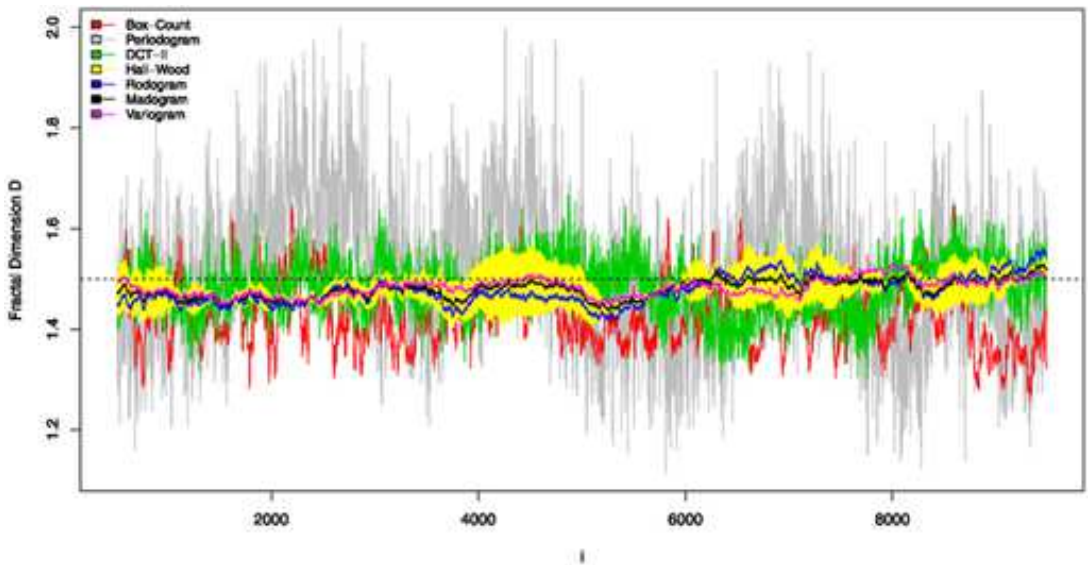}

\caption{Estimates of the fractal dimension plotted at the midpoints
of a sliding estimation window of size $1{,}025$, for a~Gaussian
sample path of the form (\protect\ref{eqtsdata}), where $n = 10{,}000$, with
exponential covariance function, $\sigma(t) = \exp(-|t|)$. The true
fractal dimension, $D = 1.5$, is marked by the dashed line. The
label on the horizontal axis, $i$, indicates the midpoint of the
sliding estimation block, at $i/10{,}000$.}
\label{figsw-1025}
\end{figure*}
%

Figure \ref{figsw-1025} illustrates these results in a further
experiment, in which we consider a Gaussian sample path with the
exponential covariance function, $\sigma(t) = \exp(-|t|)$, and
estimate the fractal dimension along sliding blocks of size $1{,}024$.
The corresponding estimates are plotted at the midpoint of the sliding
block. It is clearly seen that the variogram and other variation estimators
are the least dispersed, and that the madogram estimator is a more
efficient version of the Hall--Wood estimator. The spectral estimators
are the most dispersed, with the DCT-II estimator outperforming the
periodogram estimator, and the box-count estimator is biased.

Thus far in this section, we have studied the efficiency of the
estimators under an ideal Gaussian process assumption. However,
robustness against deviations from Gaussianity is a critically
important requirement on any practically useful estimator of the
fractal dimension. In this light, we now expand our simulation study,
and consider a situation in which Gaussian sample paths are
contaminated by additive outliers. Specifically, given a sample path
of the form (\ref{eqtsdata}), we let $i$ be discrete uniform on $\{
0, 1, \ldots, n \}$ and replace $X_{i/n}$ by $X_{i/n} + y$, where~$y$
is normal with mean zero and standard deviation 0.1 and independent of
$i$. This process is repeated to obtain the desired number of
outliers.

%
\begin{figure*}

\includegraphics{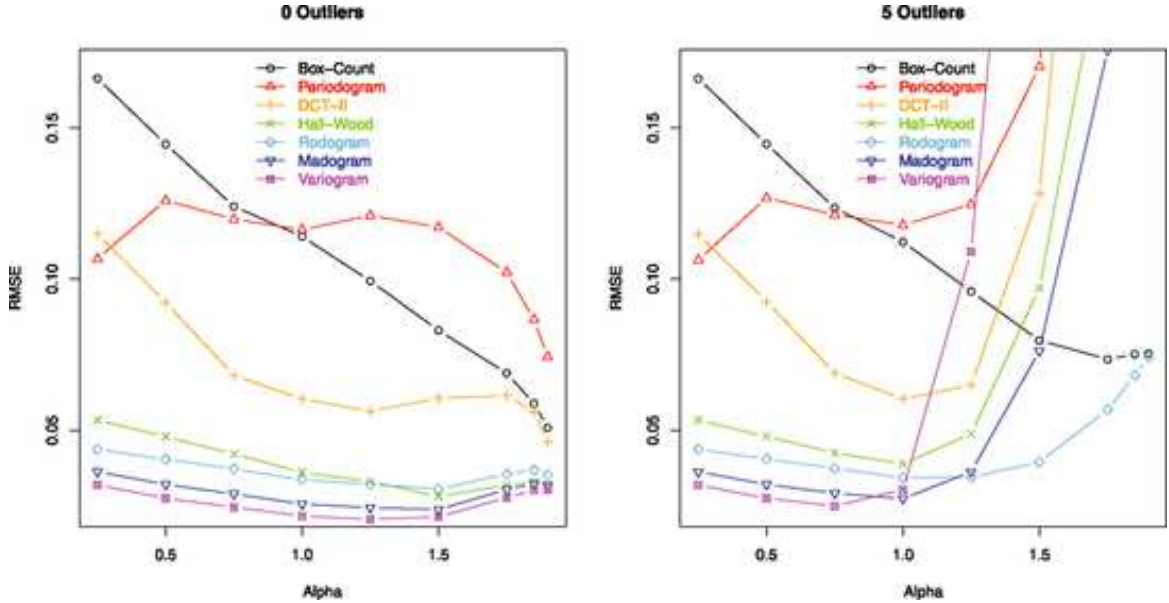}

\caption{Root mean squared error (RMSE) of estimators of fractal
dimension in dependence on the fractal index, $\alpha$, for Gaussian
sample paths of the form (\protect\ref{eqtsdata}) with sample size $n =
1{,}024$ and powered exponential covariance function, $\sigma(t) =
\exp(-|t|^\alpha)$. The panel on the left corresponds to the ideal
Gaussian process setting; the panel on the right to a situation with
five additive outliers in each sample path. The number of Monte
Carlo replicates is 1,000.} \label{fignHE-outl-alpha}\vspace*{6pt}
\end{figure*}
%

Figure \ref{fignHE-outl-alpha} shows RMSEs from such an experiment,
using sample size $n = 1{,}024$, five additive outliers, and the
powered exponential covariance function, $\sigma(t) =
\exp(-|t|^\alpha)$, with values of the fractal index~$\alpha$ that
nearly span the full range from 0 to 2. Not surprisingly, the results
resemble those in Figures \ref{figHE-p-alpha} and
\ref{figHE-outl-p-alpha}, which considered variation estimators and
the case of a single outlier only, and echo the findings of
Achard and Coeurjolly (\citeyear{achard10}). Amongst the variation
estimators considered here,\vadjust{\goodbreak}
the most outlier resistant is the rodogram estimator ($p=1/2$), but
the box-count estimator, which performs poorly overall when there are
no outliers, becomes competitive at the highest $\alpha$ considered
($1.9$).

\subsection{Discussion} \label{secdiscussion1d}

The foregoing results and arguments lead us to a recommendation for
practitioners, in that we join Bruno and Raspa (\citeyear{BR89}) and Bez and Ber\-trand (\citeyear{BB10}) and call
for the use of the madogram estimator, that is, the variation
estimator with power index $p = 1$. The madogram estimator can be
interpreted as a~statistically superior version of the Hall--Wood
estimator, is simultaneously more outlier resistant and more efficient
than many of its competitors, and has strong intuitive appeal.
Importantly, the critical relationship (\ref{eqDalpha}) between the
fractal index of a~stochastic process whose variogram of order~$p$
shows a behavior of the form (\ref{eqgammap}) at the origin, and the
fractal dimension of its sample paths, is apparently valid for a
larger class of non-Gaussian processes when $p=1$ than for certain
other choices of~$p$ (in particular, $p=2$), thereby justifying the
use of the madogram estimator for both Gaussian and non-Gaussian
stochastic processes. Its resistance to outliers can be enhanced
further if the arithmetic mean in the definition of the power
variation (\ref{eqMoM-variation}) is replaced by a~trimmed mean, as
proposed by Coeurjolly (\citeyear{coeurjolly08}).

\section{Estimating the Fractal Dimension of Spatial Data} \label{secestimators2d}\vspace*{-3pt}

We now turn to estimators of the fractal dimension of spatial data, as
discussed by Dubuc et~al. (\citeyear{dubuc89}), Constantine and Hall (\citeyear{consthall94}), Davies and Hall (\citeyear{DH99}),
Chan and Wood (\citeyear{chanwood00}) and \citet{zhustein02}, among other authors.
Burrough (\citeyear{burrough81}) noted a wealth of applications to landscape and
other environmental data, with those of Rothrock and\break Thorndike (\citeyear{rothrock80}) on the
underside of sea ice, and Goff and Jordan (\citeyear{goffjordan88}) on the topography of the
sea floor, being particularly interesting examples.

From a probabilistic point of view, a natural initial question is for
the theoretical relationship between the fractal dimension of a
surface indexed in~$\mathbb{R}^2$, and the fractal dimension of its
sections or line transects. Assuming stationarity of the spatial
random field and additional (weak) regularity conditions, Hall and Davies
(\citeyear{HD95})
showed that the fractal dimensions along line transects are all
identical to one another, except that in one special direction the
dimension may be less than in all others. Very general results that
do not depend on the stochastic process setting are available from the
fundamental work of Marstrand (\citeyear{Marstrand54}). We join Davies and Hall (\citeyear{DH99}) in
arguing that these results provide substantial support for the use of
fractal dimension as a~canonical measure of surface roughness. In
particular, they allow us to estimate the fractal dimension of a
surface by adding 1 to any estimate of the fractal dimension of the
corresponding line transects.\looseness=1

Technically, we focus discussion on the situation in which a spatial
stochastic process, indexed by the unit square in $\mathbb{R}^2$, is
sampled on a regular lattice, to yield a surface graph of the form
\begin{eqnarray} \label{eqspatialdata}
\qquad\ &&\left\{ (t, X_t)  \dvtx
t = \pmatrix{ t_1 \cr t_2
}
= \frac{1}{n} \pmatrix{ i_1 \cr i_2
} ,i_1 = 0, 1, \ldots, n, \right.\nonumber
\\[-8pt]\\[-8pt]
&&{}\hspace*{39pt}\left.\vphantom{\pmatrix{ t_1 \cr t_2}}
 i_2 = 0, 1, \ldots, n
\right\} \subset\mathbb{R}^3.\nonumber
\end{eqnarray}
Before reviewing estimators of fractal dimension studied in the extant
literature, and introducing new estimators, we propose a simple,
unified notation. Specifically, for $k > 0$ we let
\begin{eqnarray*}
S(k) &=& \left\{ (i_1,i_2,j_1,j_2) \in\{ 0, 1, \ldots, n \}^4
\dvtx\right.\vphantom{\pmatrix{ i_1 \cr i_2}}
\\
&&\hspace*{44pt}{}\left.\hspace*{3pt}\left|
\pmatrix{ i_1 \cr i_2
}
- \pmatrix{ j_1 \cr j_2
} \right|
= k \right\}  ,
\end{eqnarray*}
and denote the cardinality of this set by $N(k)$. If $N(k) > 1$, we
refer to $k$ as a relevant distance. The estimators\vadjust{\goodbreak} in the subsequent
Sections \ref{secisotropic} to \ref{secfilter} then take the
form
\begin{eqnarray} \label{eqV2d}
\widehat{D} &=& 2 - \frac{1}{p}   \biggl\{ \sum_{k  \in
\mathcal{
K}} (s_k - \bar{s}) \log\widehat{V}_p(k/n) \biggr\}\nonumber
\\[-8pt]\\[-8pt]
&&\hspace*{17pt}{}\cdot \biggl\{
\sum_{k  \in \mathcal{K}} (s_k - \bar{s})^2
\biggr\}^{-1},\nonumber
\end{eqnarray}
where $\mathcal{K}$ is a finite collection of relevant distances, $s_k =
\log(k/n)$, $\bar{s}$ is the mean of $\{ s_k \dvtx k \in\mathcal{K} \}$, and
$\widehat{V}_p(k/n)$ is a certain variation with general power index
$p > 0$.

Two-dimensional geometry allows for many options in the choice of the
distance representatives and the variation, and we restrict attention
to the most plausible and best performing estimators in the
literature, all of which are based on power variations
 (Davies and Hall, \citeyear{DH99}; Chan and Wood, \citeyear{chanwood00}; \cite{zhustein02}). As concerns the set $\mathcal{K}$ of distance
representatives, simulation experiments, experience in the line
transect case, and the work of Chan and Wood (\citeyear{chanwood00}) and \citet{zhustein02}
all suggest that a restriction to the smallest relevant distances only
tends to lead to the best performance. In addition to minimizing the
bias of the estimator, this strategy keeps the computational
complexity low as well.

\subsection{Isotropic Estimator} \label{secisotropic}

Davies and Hall (\citeyear{DH99}) considered an estimator based on the isotropic empirical
variogram, which we now generalize. For a relevant distance $k$,
consider the variation
\begin{eqnarray} \label{eqISO}
\quad \widehat{V}_{ {\rm ISO};   p}(k/n)
&=& \frac{1}{2  N(k)}\nonumber
\\[-8pt]\\[-8pt]
&&{}\cdot  \sum_{S(k)} |X_{i_1/n,   i_2/n} -
X_{j_1/n,   j_2/n}|^{ p}\nonumber
\end{eqnarray}
with general power index $p > 0$. The isotropic estimator
$\widehat{D}_{ {\rm ISO};  p}$ with power
index $p$ then is
defined by~(\ref{eqV2d}) with the set $\mathcal{K} = \{ 1, \sqrt{2}, 2
\}$ of distance representatives and the variation $\widehat{V}$ given
by (\ref{eqISO}). Thus, we consider variations at horizontal and
vertical\break spacings of one and two grid points ($k = 1$ and \mbox{$k = 2$}),
and a diagonal spacing of a single grid point ($k = \sqrt{2}$),
respectively.

\subsection{Filter Estimator} \label{secfilter}

\citet{zhustein02} studied a broad range of increment-based estimators,
among which the ``Filter~1'' estimator shows good performance. We
generalize by defining a filter estimator with general power index $p
> 0$, rather than just $p = 2$ as in\vadjust{\goodbreak} the work of \citet{zhustein02}.
Specifically, for a relevant distance $k > 0$ let
\begin{eqnarray} \label{eqF}
\widehat{V}_{ {\rm F};  p}(k/n) &=& \frac
{1}{2  N(k)}\nonumber
\\
&&{}\cdot \sum_{S(k)}
\bigl| X_{i_1/n,   i_2/n}\nonumber
\\[-8pt]\\[-8pt]
&&{}\hspace*{25pt}- 2  X_{(i_1+j_1)/(2n),   (i_2+j_2)/(2n)}\nonumber
\\
&&{}\hspace*{63pt}\hspace*{25pt}+ X_{j_1/n,   j_2/n}
\bigr|^{ p}.\nonumber
\end{eqnarray}
The filter estimator $\widehat{D}_{ {\rm F}; \hspace
{0.3mm}p}$ with power
index $p$ then is defined by (\ref{eqV2d}) with the set $\mathcal{K} =
\{ 2, 2 \sqrt{2}, 4 \}$ of distance representatives and the variation
$\widehat{V}$ given by (\ref{eqF}). This considers variations at
horizontal and vertical spacings of one and two grid points (\mbox{$k = 1$}
and $k = 2$), and a diagonal spacing of a single grid point ($k =
\sqrt{2}$), respectively. Hence, the filter estimator is the natural
equivalent of the isotropic estimator $\widehat{D}_{{\rm ISO};
p}$, but now using second differences, rather than first
differences.

\subsection{Square Increment Estimator} \label{secsquare-increment}

The square increment estimator is based on a proposal of
Chan and Wood (\citeyear{chanwood00}), who restricted\break attention to quadratic variations.
Here we define\break a~square increment variation of general power index $p
> 0$, namely,
%
\begin{eqnarray} \label{eqSQ}
\widehat{V}_{ {\rm SI};  p}(k/n) &= &\frac
{1}{2  N(k)}\nonumber
\\
&&{}\cdot \sum_{S(k)}
| X_{i_1/n,   i_2/n}- X_{i_1/n,   j_2/n}
\\
&&{}\hspace*{25pt}- X_{j_1/n,   i_2/n} + X_{j_1/n,   j_2/n}
|^{ p},\nonumber
\end{eqnarray}
where $k$ is a relevant distance. The square increment estimator
$\widehat{D}_{ {\rm SI};  p}$ then is
defined by
(\ref{eqV2d}) with the set $\mathcal{K} = \{ \sqrt{2}, 2 \sqrt{2} \}$ of
distance representatives, corresponding to squares that have side
widths of one and two grid points, and the variation $\widehat{V}$
given by (\ref{eqSQ}).

\subsection{Transect Estimators} \label{secsimple}

Finally, we consider two very simple estimators that are based on the
variation estimator $\widehat{D}_{{\mathrm V}; p}$ with general power
index $p > 0$ of Section
\ref{secvariation}, or a variant that uses second differences, as
defined in equation~(\ref{eqMoM-variation-1}). In either case, a line
transect estimate of the fractal dimension is computed for each row and
each column in the grid. The transect-variation and
transect-increment estimators $\widehat{D}_{ {\rm TV};
 p}$
and $\widehat{D}_{ {\rm TI};  p}$ with
power index $p$ then
add 1 to the median of the $2n$ corresponding line transect estimates.
In the former\vadjust{\goodbreak} case, the line transect estimates are based on first
differences, in the latter on second differences.

\section{Performance Assessment: Spatial Data} \label{secperformance2d}

We now assess the estimators by their large sample behavior under
infill asymptotic as well as in finite sample simulation studies,
considering both efficiency and robustness.

\begin{figure*}

\includegraphics{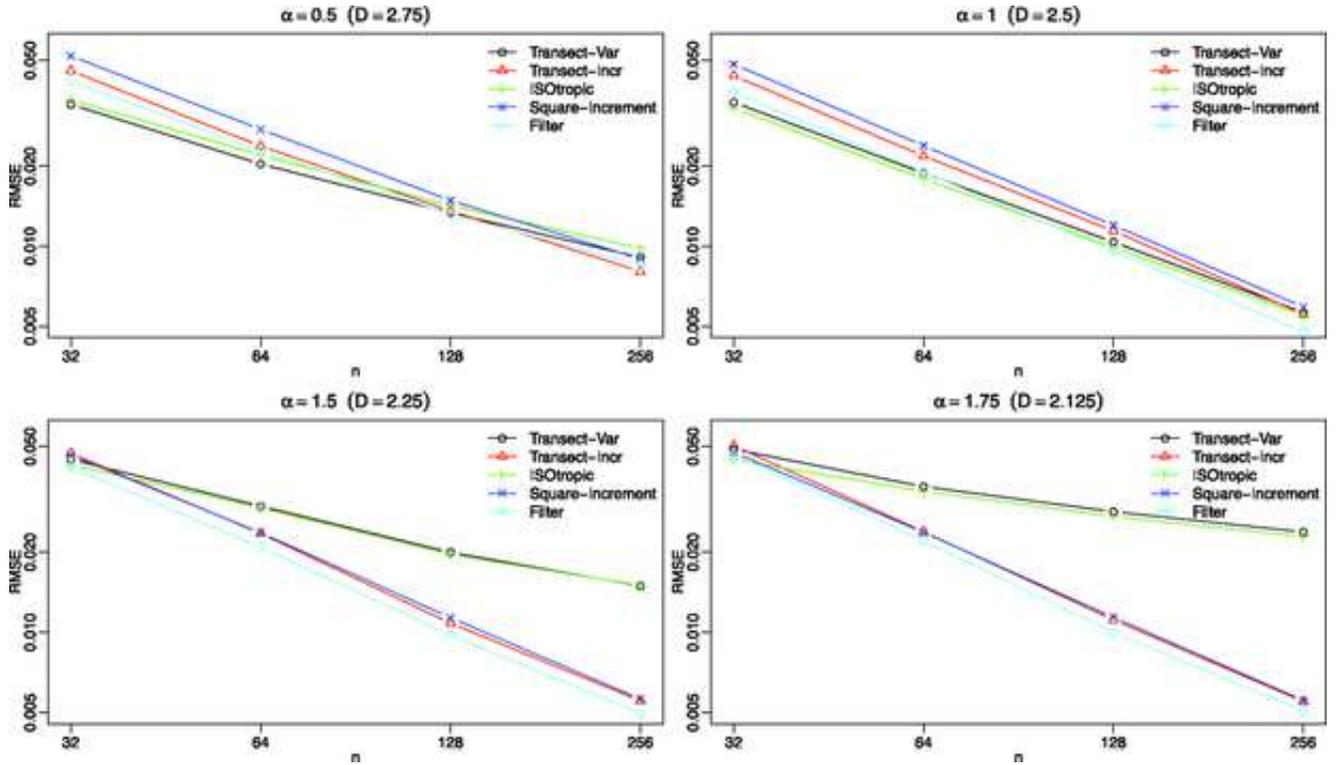}

\caption{Root mean squared error (RMSE) for estimators of fractal
dimension for data of the form (\protect\ref{eqspatialdata}) from spatial
stochastic processes with powered exponential covariance, $\sigma(t)
= \exp(-|t|^\alpha)$, versus the (square root of the)
sample size, $n$. The power index used is $p = 2$, and for each
combination of $\alpha$ and $n$, the number of Monte Carlo
replicates is 1,000.} \label{figrmse2d-p=2}\vspace*{6pt}
\end{figure*}

\subsection{Asymptotic Theory} \label{secasymptotic2d}

Davies and Hall (\citeyear{DH99}),  Chan~and Wood (\citeyear{chanwood00})~and \citet{zhustein02} developed
asymptotic theory for a very wide range of estimators of the form~(\ref{eqV2d}) that are based on variations with power index $p = 2$.
Generally, their results apply under an infill asymptotic scenario for
sample paths of the form~(\ref{eqspatialdata}) from an intrinsically
stationary Gaussian spatial process with fractal index $\alpha\in
(0,2)$ and a variogram that behaves like~(\ref{eqvariogram}) at the
origin. For estimators that are based on variations corresponding to
a first difference, the generic result is that
\begin{eqnarray} \label{eqasymptoticrate2d}
&&\mbox{MSE}(\widehat{D})\nonumber
\\[-8pt]\\[-8pt]
&&\quad =
\cases{
\mathcal{O}(n^{-2}) + \mathcal{O}(n^{-4\beta}), & \mbox{if }$  0 < \alpha< 1$,
\cr
\mathcal{O}(n^{-2} L(n)) + \mathcal{O}(n^{-4\beta}), & \mbox{if }$ \alpha= 1$,
 \cr
\mathcal{O}(n^{2\alpha-4}) + \mathcal{O}(n^{-4\beta}), & \mbox{if }$  1 <
\alpha< 2$,
}\nonumber\hspace*{-25pt}
\end{eqnarray}
where $L$ is a function which is slowly varying at infinity. If
$\alpha\leq1$, then $\widehat{D}$ has a normal limit, while, if
$\alpha> 1$, the limit is related to a Rosenblatt distribution, with
some of these results carrying over to certain specific non-Gaussian
processes (Chan and Wood, \citeyear{chanwood04}). However, if the variations correspond
to a second difference, such as in the cases of the filter and square
increment estimators, and/or the generalized least squares techniques,
rather than the ordinary least squares method, is used, an improved
asymptotic behavior, namely,
%
\begin{equation} \label{eqgoodasymptoticrate2d}
 \mbox{MSE}(\widehat{D}) = \mathcal{O}(n^{-2})+ \mathcal{O}(n^{-4\beta})
\end{equation}
for all $0 < \alpha< 2$, is achieved, with an associated limit distribution that is normal.
For regularity conditions and further detail, we refer to the original
work of Davies and Hall (\citeyear{DH99}), Chan and Wood (\citeyear{chanwood00}), \citet{zhustein02} and
Chan and Wood (\citeyear{chanwood04}), which is impressive. While these authors restricted
attention to quadratic variations with power index $p = 2$ only, we
conjecture, based on the work of Guyon and Le{\'o}n (\citeyear{guyonleon89}),
Istas and Lang (\citeyear{istaslang97}) and Barndorff-Nielsen, Corcuera
and Podolskij (\citeyear{barndorff-nielsen09}) on the power\vadjust{\goodbreak}
variations of Gaussian processes, that analogous results continue to
hold under a general power index $p > 0$, similar to the line transect
case studied by Coeurjolly (\citeyear{coeurjolly08}).

As the amount of data in the sampling scheme~(\ref{eqspatialdata}) is
about $n^2$, the asymptotic rates of convergence in~(\ref{eqasymptoticrate2d}) and (\ref{eqgoodasymptoticrate2d})
conform with those in the time series or line transect case, except
that the transition from the classical to slower rates of convergence
occurs already at $\alpha= 1$ (or $D = 3/2$), rather than at $\alpha
= 3/2$ (or $D = 7/4$). Hence, these results suggest that there may be
a higher benefit to using increments that are based on second
differences in the spatial case than in the time series or line
transect case.

\subsection{Simulation Study: Gaussian Spatial Processes} \label{secsimulation2d}

In the subsequent simulation study, we use state-of-the-art
implementations of the circulant embedding method (Stein, \citeyear{stein02}; Gneiting et~al., \citeyear{Getal06}) to generate Gaussian sample surfaces.
Figure~\ref{figrmse2d-p=2} shows the root mean squared error (RMSE)
of the estimators versus the (square root of the) sample
size,~$n$, using data of the form (\ref{eqspatialdata}) from
stationary spatial processes with powered exponential covariance
function, $\sigma(t) = \exp(-|t|^\alpha)$. The estimators use the
traditional power index $p = 2$, and the number of Monte Carlo
replicates is 1,000.

%

%

The asymptotic rates of convergence in (\ref{eqasymptoticrate2d})\break
and~(\ref{eqgoodasymptoticrate2d}) suggest linear graphs with
slope $-1$ on the logarithmic scale for the filter and square
increment estimators and all values of the fractal index, $\alpha$.
For the isotropic estimator, they suggest slope $-1$ for $\alpha\leq
1$, slope $-1/2$ for $\alpha= 3/2$, and slope $-1/4$ for $\alpha=
7/4$. Our empirical results are in good agreement with the
theoretical slopes, and attest to the superior performance of the
filter estimator in the ideal Gaussian process setting, as noted by
\citet{zhustein02}. Also, these and other simulation results lead us
to conjecture that the transect-variation estimator behaves like
(\ref{eqasymptoticrate2d}), while the transect-increment estimator
shares the favorable uniform asymptotic rate of convergence
in~(\ref{eqgoodasymptoticrate2d}).

\begin{sidewaysfigure}
\vspace*{280pt}
\centering
\includegraphics{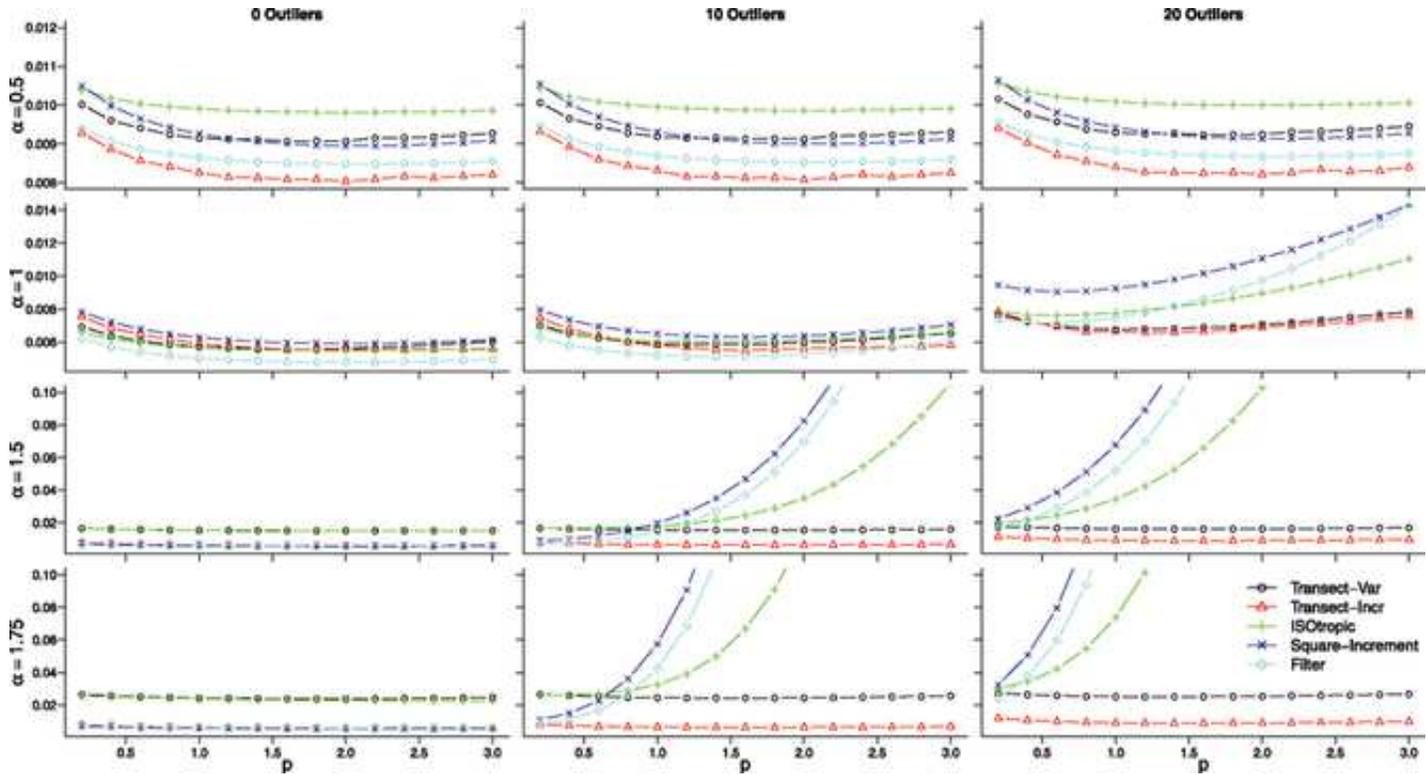}

\caption{Root mean squared error of isotropic, filter, square
increment, transect-variation and transect-increment estimators of
fractal dimension in dependence on the power index, $p$. The
underlying sample paths are of the form (\protect\ref{eqspatialdata}) with
$n = 256$ from Gaussian spatial processes with powered exponential
covariance, $\sigma(t) = \exp(-|t|^\alpha)$. The columns correspond
to situations with no outliers (left), 10 additive outliers
(middle) and 20 additive outliers (right), respectively. The rows
are for distinct values of the fractal index, namely $\alpha= 0.5,
1.0, 1.5$ and $1.75$. The number of Monte Carlo replicates is 1,000.}
\label{figrmse2d-outliers}
\end{sidewaysfigure}

In Figure \ref{figrmse2d-outliers} we consider estimators with
general power index $p > 0$, but fix $n = 256$ in the spatial sampling
scheme (\ref{eqspatialdata}). Again, we use the powered exponential
covariance model, and the number of Monte Carlo replicates is 1,000,
each comprising a~total of $(n+1)^2 = 66{,}049$ observations within the
unit square. The left column shows the RMSE in the ideal Gaussian
process setting, in which the filter,\vadjust{\goodbreak} square increment and
transect-increment estimators perform best. Furthermore, the
efficiency of these estimators depends only very little on the choice
of the power index.

Figure \ref{figrmse2d-outliers} also studies the behavior of the
estimators in the presence of outliers. Specifically, the middle and
right-hand columns show RMSEs in situations in which the Gaussian
sample paths have been contaminated by 10 and 20 additive outliers,
respectively, in ways essentially identical to those described in
Section \ref{secsimulation1d}. Note that the vertical scale differs
from row to row, with the largest RMSEs corresponding to the smoothest
surfaces. The smoother the surface, that is, the larger the value of
the fractal index, $\alpha$, the more impact the outliers have on the
estimators. Two of the estimators that dominate in the uncontaminated
setting, namely the filter and the square increment estimators, are
the least outlier resistant, even though the outlier fraction is very
small in our study, at 0.015 and 0.03 percent, respectively. The most
resistant estimators are the transect estimators.

These results allow for an interpretation in terms of breakdown
points. While comprehensive formal definitions of breakdown points
for dependent data have recently become available  (Genton and Lucas, \citeyear{GL05}), it
suffices here to take a heuristic point of view, and define the
breakdown point of an estimator as the minimal fraction of
contaminated data that can ruin an estimator. Evidently, the
isotropic, filter and square increment estimators have breakdown point
zero. In contrast, the transect estimators have a positive breakdown
point of about $1/(2  \sqrt{m})$ under the spatial
sampling scheme
(\ref{eqspatialdata}), where \mbox{$m\,{=}\,n^2$} is the approximate amount of
data. To see this, note that each outlier affects the individual
estimate for at most two of the $2  n$ transects, from
which the
median is formed. Thus, a transect estimator cannot be ruined, unless
a fraction of at least $(n/2)/n^2$, or $1/(2  \sqrt
{m})$, of
the data are contaminated.\vspace*{2pt}

\subsection{Discussion} \label{secdiscussion2d}\vspace*{2pt}

While confirming extant theoretical and simulation results for spatial
data, which suggest the use of variations that are based on second
differences, the results of this section lead us to two novel
insights.

%
\clearpage
%
\clearpage

\begin{figure*}

\includegraphics{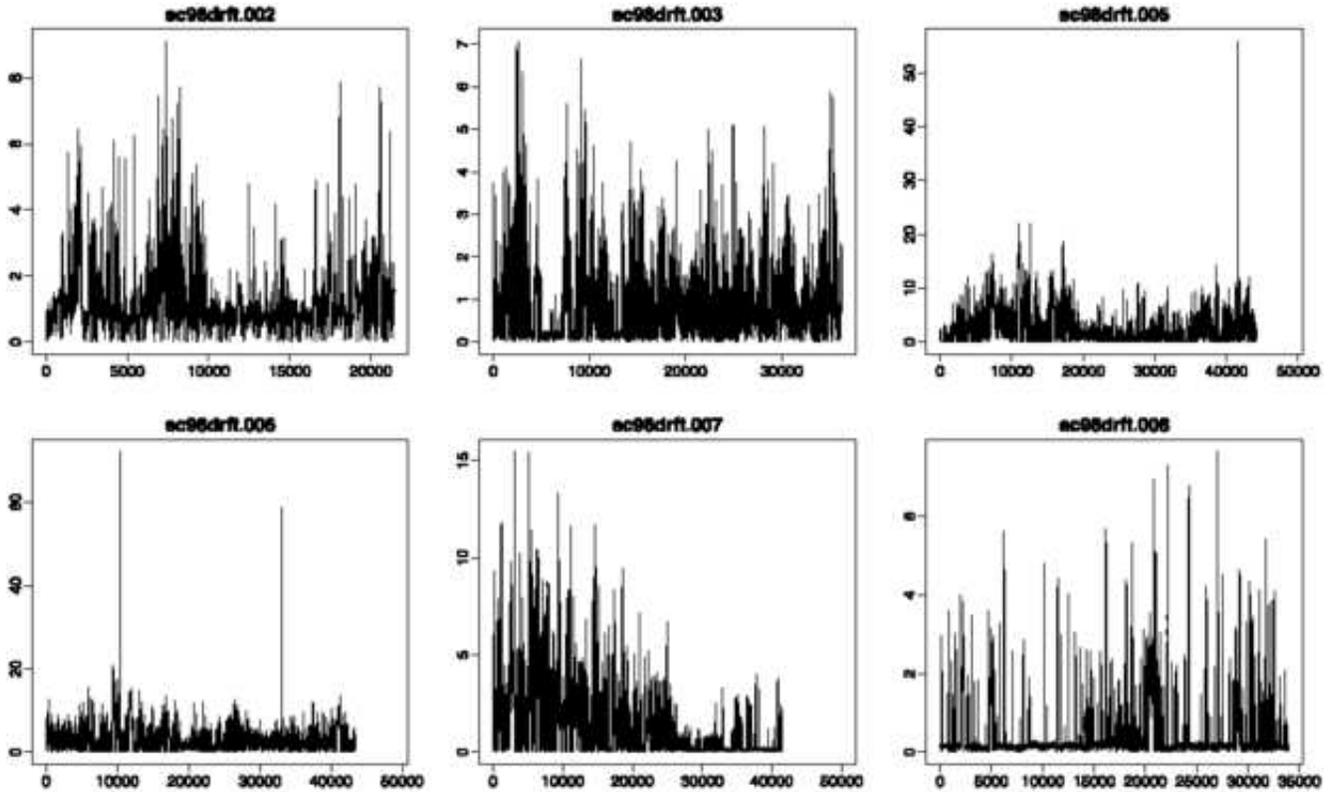}

\caption{Arctic sea-ice profiles, with the panels showing files
sc98drft.002, 003 and 005--008, respectively, and the labels being
in meters. See the text for details.} \label{figiceprofiles}
\end{figure*}

Thus far, the statistical literature has restricted attention to
variation-based estimators with power index $p = 2$. A first
observation is that the efficiency of these estimators depends little
on the power index, $p > 0$. Outlier resistance and robustness
arguments then suggest the use of smaller power indices, with Section
\ref{secvariation} supporting the choice of $p = 1$.

A second and potentially very surprising insight is the superior
performance of the transect-variation and transect-increment
estimators. These estimators have positive breakdown points and
outperform the traditional estimators even under minimal deviations
from the ideal Gaussian process setting. Indeed, in practice, we
would expect much larger deviations than in our simulation setting, in
which the outliers had substantially lower variability than the
process itself, and occurred at fractions of about 1 in 6,000 and 1 in
3,000 only. As the transect-increment estimator appears to share the
favorable uniform rate of convergence
(\ref{eqgoodasymptoticrate2d}) with the best performing extant
estimators, this suggests the availability of an estimator, namely the
transect-increment estimator with power index $p = 1$, that is both
robust and efficient. We believe that these are highly promising
prospects that deserve further study.

If spatial data are observed at irregularly scattered locations,
rather than a regular grid, the only available estimator is a suitably
modified version of the isotropic estimator (\ref{eqISO}), with
$S(k)$ now representing pairs of stations that are approximately
a~distance $k$ apart. In such cases, the use of the power index $p =
1$ again seems prudent.

\section{Data Example: Arctic Sea-Ice Profiles} \label{secdata}

%

\begin{figure*}

\includegraphics{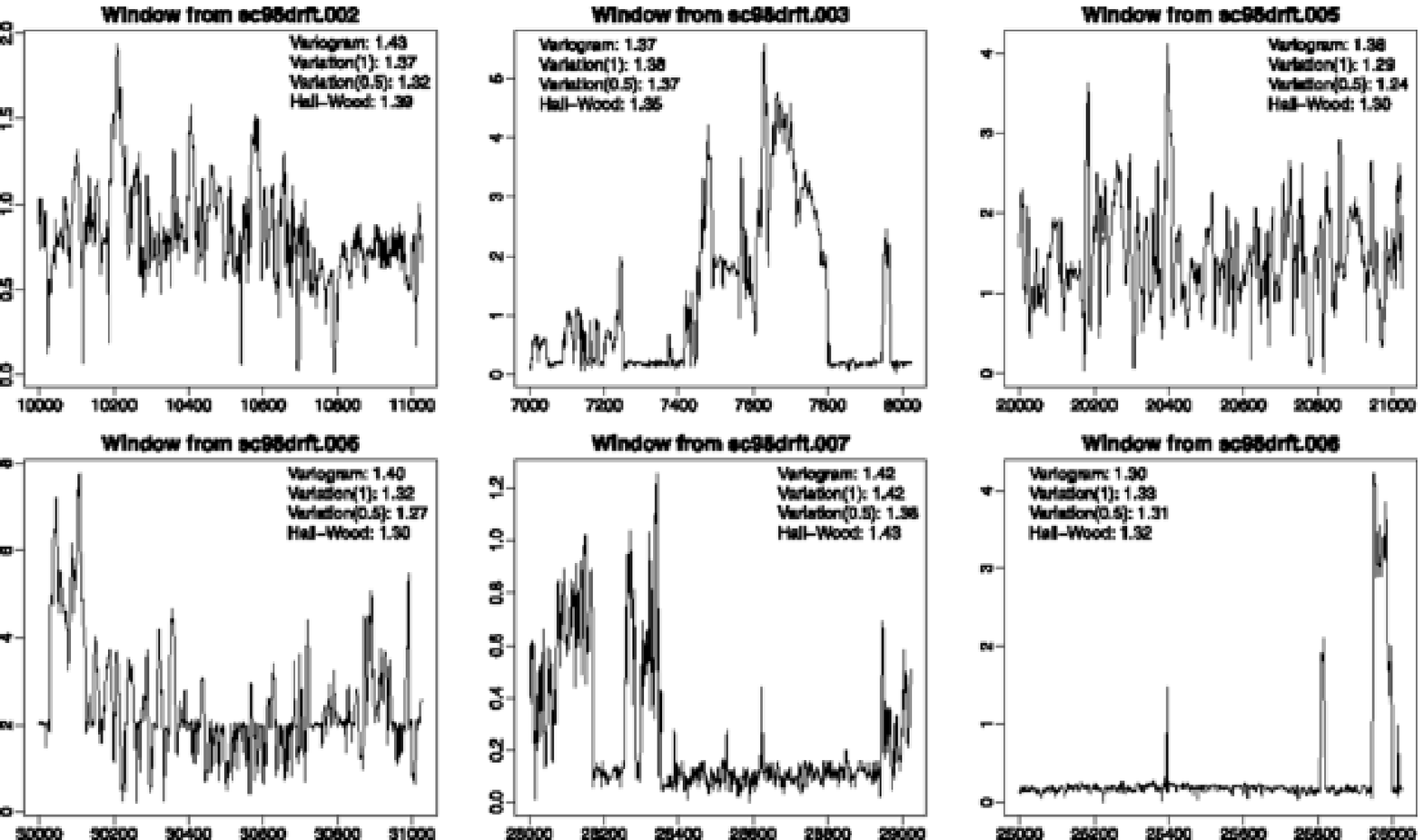}

\caption{Selected blocks of width 1,024 meters from the sea-ice profiles
in Figure \protect\ref{figiceprofiles}, along with the corresponding
estimates of fractal dimension.} \label{figiceprofiles-w}\vspace*{-4pt}
\end{figure*}

In this penultimate section, we estimate the fractal dimension of
arctic sea-ice profiles based upon measurements of sea-ice draft
(93\% of thickness). These profiles can be regarded as line
transects\break
through the underwater surface of sea ice because they were collected
using upward-looking sonars by submarines traveling under the sea ice
with no appreciable deviations from a single direction and\break depth.
The data used here were sampled one meter apart at a resolution
of one meter by the U.S.~Navy~in August 1998 in the Arctic Ocean,
and are available online from the National Snow and Ice Data~Center
at
\texttt{\href{http://nsidc.org/data/docs/noaa/g01360\_upward\_looking\_sonar}{http://nsidc.org/data/docs/noaa/g01360\_}\break
\href{http://nsidc.org/data/docs/noaa/g01360\_upward\_looking\_sonar}{upward\_looking\_sonar}}.
We examined six profiles of about 240 km total length (files sc98drft.002,
003 and 005--008). These profiles are illustrated in
Figures \ref{figiceprofiles} and\vadjust{\goodbreak} \ref{figiceprofiles-w} and show
pronounced non-Gaussian features.
Given the resolution, we join the extant literature in using fractal
dimension to characterize the surface roughness of the macroscopic,
topographic structure of sea ice, so that meters correspond
to sufficiently small scales.

\begin{figure*}[b]
\vspace*{-4pt}
\includegraphics{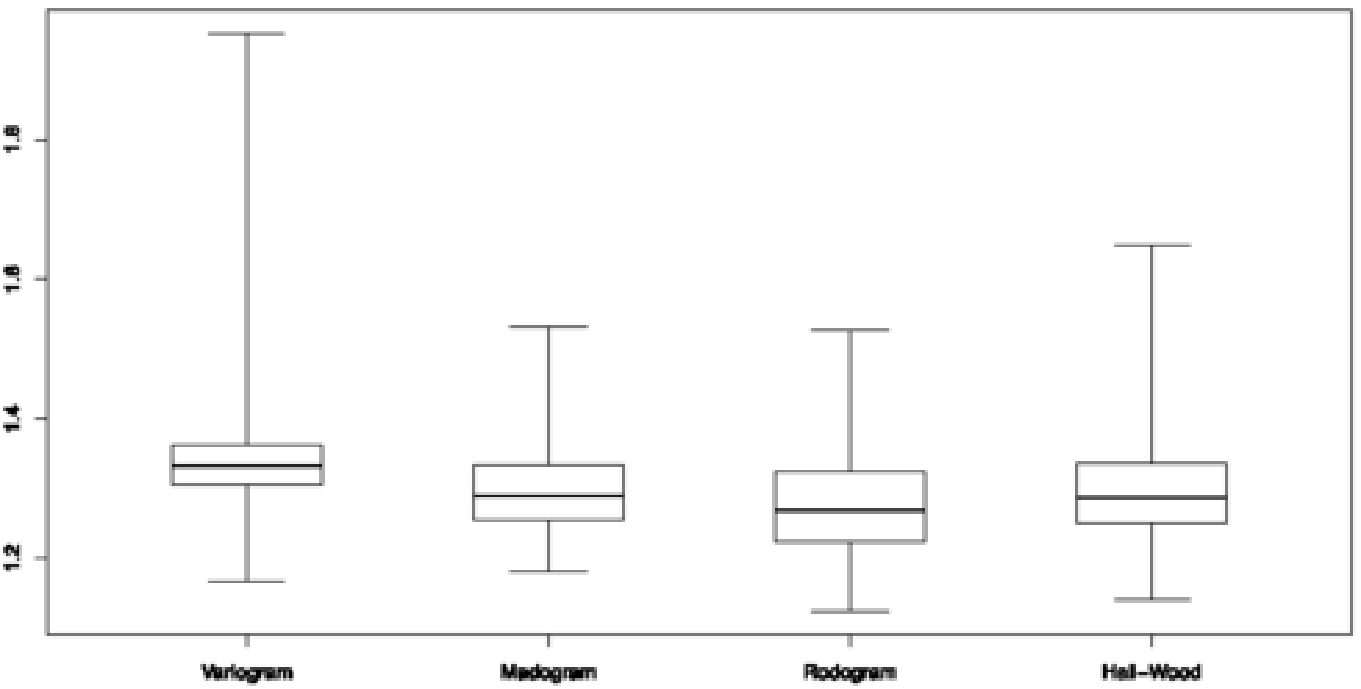}

\caption{Boxplots for variation estimates with power indices $p = 2$
(variogram),
$1$ (madogram) and $1/2$ (rodogram) along with Hall--Wood estimates of fractal
dimension for the blocks in the ice profile data. See the text for details.}
\label{figboxplot-ice}
\end{figure*}

\begin{figure*}[t]

\includegraphics{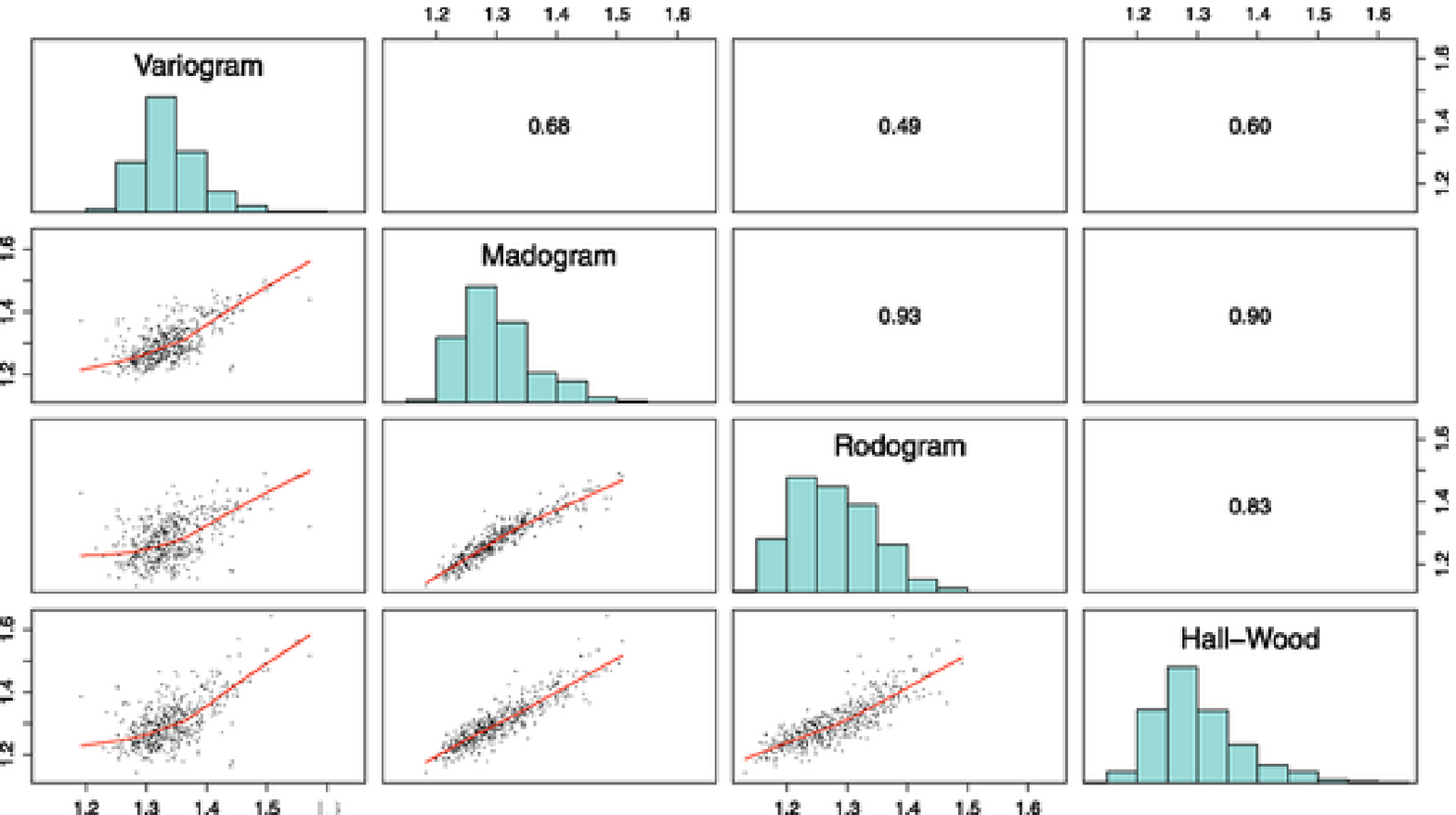}

\caption{Scatterplot matrix for variation estimates with power indices
$p = 2$ (variogram), $1$ (madogram) and $1/2$ (rodogram) along with
Hall--Wood estimates of
fractal dimension for the blocks in the ice profile data. See the text
for details.} \label{figscatterplot-matrix-ice}\vspace*{3pt}
\end{figure*}

We use a sliding estimation window of width $n = 1{,}024$ meters, move
these blocks along the profiles in increments of 10 meters, and
estimate fractal dimension for each of them, using variation estimators
with power indices $p = 2$ (variogram), $1$ (madogram) and
$1/2$ (rodogram) along with the Hall--Wood estimator. Thus, for each method
there are 23,303
blocks in total, with Figure \ref{figiceprofiles-w} showing examples
of profiles and dimension estimates. Figures \ref{figboxplot-ice}
and \ref{figscatterplot-matrix-ice} show the corresponding boxplots,
histograms and pairwise scatter diagrams, composited over all blocks.
While all four methods result in similar estimates, the strongest
correlation is, not surprisingly, between the Hall--Wood and madogram
estimators, with the latter being our estimator of choice.

%

%

%

Overall, the Arctic sea-ice profiles have Hausdorff dimension of
about 1.3 along line transects, and thus of about 2.3 spatially.
These estimates are at the lower end of results reported in the
literature. For instance, Bishop and Chellis (\citeyear{bishop89}) argued that profiles
of ice keels have fractal dimensions ranging from 1.2 to 1.7, while
Connors, Levine and Shell (\citeyear{connors90}) reported estimates of about 1.4 and 1.6 for first
year and multiyear ice segments, respectively. Further results and
background information can be found in the works of Rothrock and Thorndike
(\citeyear{rothrock80})
and Goff et~al. (\citeyear{goff95b}).

\section{Concluding Remarks} \label{secdiscussion}

In closing this review, we return to the opening quote of
 Lenny Smith [(\citeyear{Smith07}), page~115] that precedes the abstract
of our paper. Indeed, healthy skepticism about dimension estimates
based on real-world data is well justified, in that estimates of
fractal dimension depend on the availability of data at suitable
scales and face issues of discreteness, and the effects of measurement
error might also need to be disentangled. Notwithstanding these
issues, estimates of fractal dimension can serve as informative
descriptors of surface roughness.

In the case of time series or line transect data, we recommend the use
of the madogram estimator, that is, the variation estimator with power
index $p = 1$. In the case of spatial lattice data, transect
estimators based on madogram estimators along rows and columns show
high promise. These recommendations\vadjust{\goodbreak} echo observations of Bruno and Raspa
(\citeyear{BR89})
and Bez and Ber\-trand (\citeyear{BB10}), who argued that the critical relationship
(\ref{eqDalpha}) between the fractal index of a stochastic process
whose variogram of order $p > 0$ shows a behavior of the form
(\ref{eqgammap}) at the origin, and the fractal dimension of its
sample paths, is particularly robust and inclusive when $p = 1$. We
encourage work toward rigorous results in these directions, including
both variograms and their equivalents for general types of line
transect and spatial increments.

Furthermore, we call for the development of large sample theory for
variation estimators of general power index, including but not limited
to our preferred choice of $p = 1$. While for time series this has
been achieved in the far-reaching recent work of Coeurjolly (\citeyear{coeurjolly08}),
the case of spatial data remains open. As noted in Section
\ref{secasymptotic2d}, we believe that many of the results in the
extant default case $p = 2$ can be carried over to a general power
index, $p > 0$. In doing so, the results of Guyon and Le{\'o}n (\citeyear{guyonleon89}),
Istas and Lang (\citeyear{istaslang97}) and Barndorff-Nielsen, Corcuera and Podolskij (\citeyear{barndorff-nielsen09}) on the power
variations of Gaussian processes provide tools that can be applied in
concert with the methodology put forth in an impressive\vadjust{\goodbreak} strand of
asymptotic literature for $p = 2$, which includes the work of
Hall and Wood (\citeyear{hallwood93}), Constantine and Hall (\citeyear{consthall94}), Kent and Wood (\citeyear{kentwood97}), Davies and Hall (\citeyear{DH99}),
Chan and Wood (\citeyear{chanwood00}) and \citet{zhustein02}, among others. Of particular
interest is our conjecture in the spatial setting of Section
\ref{secasymptotic2d}, according to which the transect-variation
approach allows for estimators of the fractal dimension that are
simultaneously highly efficient and robust. This approach can be
paired with the use of trimmed means or linear combinations of sample
quantiles in defining power variations, as suggested by
Coeurjolly (\citeyear{coeurjolly08}) for time series, or with the use of bipower and
multipower variations (Barndorff-Nielsen et~al. \citeyear{BGJPS2006}; Barndorff-Nielsen, Corcuera and Podolskij, \citeyear{CBNP2011}). While the case of
Gaussian processes appears to be challenging yet tractable, a general
asymptotic theory for non-Gaussian processes remains elusive, despite
the recent progress by Chan and Wood (\citeyear{chanwood04}) and Achard and Coeurjolly (\citeyear{achard10}).

To our knowledge, Bayesian methods for estimating fractal dimension
have not been explored yet, except that the method of Handcock and Stein
(\citeyear{HS93})
could be applied in a parametric context. Physical insight can drive
the choice of the prior, and the development of Bayesian madogram
estimators might be a promising option.

Multifractional Brownian motion (Peltier and Levy Vehel, \citeyear{peltier95};\vadjust{\goodbreak} Benassi, Jaffard and
Roux,
\citeyear{benassi97};
Herbin, \citeyear{herbin06})
is a generalization of the classical fractional Brownian motion, where
the fractal dimension is allowed to vary along the sample paths.
Similarly, the nonstationary Gaussian random fields described by
Anderes and Stein (\citeyear{anderesstein10}) allow for location-depend\-ent, local Hausdorff
dimensions, which need to be estimated as functions, rather than a
single number. In this setting, the estimators considered in our
paper can be used as building blocks for the more complex estimators
needed to handle these nonstationary processes,
as studied by Benassi, Cohen and Istas (\citeyear{benassi98}), Ayache and L{\'e}vy~V{\'e}hel (\citeyear{ayache04}) and Coeurjolly (\citeyear{coeurjolly05}),
among others. For an applied perspective, see Gagnon, Lovejoy and Schertzer
(\citeyear{gagnonetal06}).

To close on a practical note, we have developed an \texttt{R} package,
called \texttt{fractaldim}, that implements the estimators of fractal
dimension discussed in this paper ({\v{S}}ev{\v{c}}{\'{\i}}kov{\'a}, Gneiting and Percival, \citeyear{sevc10}). It has the ability
to compute estimates for a single dataset, or a series of
estimates along sliding blocks, as in our data example. Multiple
estimates can be conveniently bundled into a single call, and the
default arguments correspond to the recommendations in this paper.

For example, to generate four log--log plots of the type shown in
Figure \ref{figbox-count}, a few lines of code suffice:\vspace*{6pt}
\begin{verbatim}
par(mfrow=c(2,2))
series <- GaussRF(x=c(0,1,1/1024),
  model='stable', gridtriple=TRUE,
  param=c(mean=0, variance=1,
  scale=1, kappa=1))
methods <- c('hallwood',
  'periodogram', 'variogram',
  'madogram'))
D <- fd.estimate(series,
  method=methods, plot.loglog=T,
  plot.allpoints=T)
\end{verbatim}\vspace*{6pt}
Here, the function \texttt{GaussRF} from the \texttt{Random\-Fields} package
is used to generate a Gaussian sample path with exponential covariance
and fractal index $\alpha= 1$. The desired estimation methods are
passed to the estimation function, \texttt{fd.estimate}, as character
strings as in our example. Alternatively, each method can be given as a list
with a \texttt{name} element
determining the method, and any additional elements specifying
parameters. For example, an entry \texttt{list(name='variation',
p.index=1)} is equivalent
to \texttt{'madogram'}.

To obtain interval estimates with a parametric bootstrap method, as
proposed by Davies and Hall (\citeyear{DH99}) and shown in many of our figures, the
following call suffices, using the simulated \texttt{series} from above,
\texttt{boot.it} bootstrap replicates, and the madogram estimator:\vspace*{6pt}
\begin{verbatim}
D <- fd.estimate(series,
  methods='madogram')
D.boot <- rep(NA, boot.it)
alpha <- 4 - 2*D$fd
for (i in 1:boot.it)
 {
 boot.series <- GaussRF(
   x=c(0,1,1/1024), model='stable',
   gridtriple=TRUE,
   param=c(mean=0, variance=1,
   scale=D$scale^(-1/alpha),
   kappa=alpha))
 D.boot[i] <- fd.estimate(
   boot.series,
   methods='madogram')$fd
 }
\end{verbatim}\vspace*{6pt}
Confidence intervals can be obtained from the array \texttt{D.boot} in
the usual way. Additional functions for variation estimators in the
line transect case are available within the {\sc dvfBm} package
(Coeurjolly,\break \citeyear{dvfBm}).

\section*{Acknowledgments}

We are grateful to two anonymous reviewers, Nicolas Bez, Werner Ehm,
Marc Genton, Mark Podolskij, Roopesh Ranjan, Michael Scheuerer and
Michael Zaiser for comments and discussions. Our research has been
supported by the National Science Foundation under Awards DMS-0706745
and ARC-0529955, by the Alfried Krupp von Bohlen und Halbach
Foundation and by Grant Number R01 HD054511 from the National
Institute of Child Health and Human Development.


\end{document}